\newif\iffigs\figstrue
\documentclass[12pt]{article}
\usepackage{latexsym,amssymb}
   \input{epsf}
\else
\fi

\newcommand{\sect}[1]{\setcounter{equation}{0}\section{#1}}

\newcommand{\app}[1]{\setcounter{section}{0}\setcounter{equation}{0}
\renewcommand{\thesection}{\Alph{section}}
\section{#1}}

\newcommand{\eq}{\begin{equation}}
\newcommand{\eqa}{\begin{eqnarray}}
\newcommand{\en}{\end{equation}}
\newcommand{\ena}{\end{eqnarray}}
\newcommand{\enn}{\nonumber \end{equation}}

\def\sk{\vskip .4cm}
\def\noi{\noindent}

\def\al{\alpha}

\def\be{\beta}
\def\ga{\gamma}
\def\Ga{\Gamma}

\def\Cb{\bar{C}}

\def\rhop{\rho'}

\def\epsi{\varepsilon}
\def\we{\wedge}
\def\th{\theta}
\def\de{\delta}

\def\part{\partial}

\def\L#1#2{ \La^{#1}_{~~~#2} }

\def\La{\Lambda}

\def\Cb{{\bf \mbox{\boldmath $C$}}}

\def\DR{\Delta_R}
\def\DL{\Delta_L}

\def\D{\Delta}

\def\n2{{{N+1} \over 2}}

\def\square{{\,\lower0.9pt\vbox{\hrule \hbox{\vrule height 0.2 cm
\hskip 0.2 cm \vrule height 0.2 cm}\hrule}\,}}

\def\Q.E.D.{\rightline{$\Box$}}

\def\sumong{\sum_{g \in G}}

\def\sumongnote{\sum_{g \not= e}}

\def\sumonh{\sum_{h \in G}}

\def\Lcal{{\cal L}}
\def\Rcal{{\cal R}}

\def\Gcal{{\cal G}}

\def\lb{\label}

\def\Ncal{{\cal N}}

\def\ell{{{t}}}

\newcommand{\matc}{\begin{array}{c}}
\newcommand{\matcc}{\begin{array}{cc}}
\newcommand{\matccc}{\begin{array}{ccc}}
\newcommand{\matcccc}{\begin{array}{cccc}}
\newcommand{\emat}{\end{array}}

\begin{document}
\begin{titlepage}
\vskip -1cm \rightline{DFTT-33/01} \rightline{LMU-TPW-01-15}
\rightline{January 2002}
 \vskip 1em
\begin{center}
{\Large\bf Discretized Yang-Mills and Born-Infeld actions\\[.2em]
on finite group geometries }
\\[2em]
{\bf P. Aschieri}${}^{1}$, {\bf L. Castellani}${}^{2}$ and {\bf
A.P. Isaev }${}^3$\\[3em] {\sl ${}^1$ Sektion Physik der
Ludwig-Maximilians-Universit\"at,\\ Theresienstr. 37, D-80333
M\"unchen, Germany}
 \\[.2em] {\sl ${}^2$ Dipartimento di Scienze,
 Universit\`a del Piemonte Orientale,\\ Dipartimento di
Fisica Teorica and I.N.F.N\\ Via P. Giuria 1, 10125 Torino,
Italy.} \\[.2em] {\sl  ${}^3$ Bogoliubov Laboratory of Theoretical
Physics, JINR, \\ 141 980 Dubna, Moscow Region, Russia}
\end{center}
 \vskip 2 cm
\begin{abstract}
Discretized nonabelian gauge theories living on finite group
spaces $G$ are defined by means of a geometric action $\int Tr~ F
\we *F$. This technique is extended to obtain discrete versions of
the Born-Infeld action. The discretizations are in 1-1
correspondence with differential calculi on finite groups.

A consistency condition for duality invariance of the discretized
field equations is derived for discretized $U(1)$ actions $S[F]$
living on a 4-dimensional abelian $G$. Discretized
electromagnetism satisfies this condition and therefore admits
duality rotations.

Yang-Mills and Born-Infeld theories are also considered on product
spaces $M^D \times G$, and we find the corresponding field
theories on $M^D$ after Kaluza-Klein reduction on the $G$ discrete
internal spaces. We examine in some detail the case $G=Z_N$, and
discuss the limit $N \rightarrow \infty$.

 A self-contained review
on the noncommutative differential geometry of finite groups is
included.

\end{abstract}

\vskip 1 cm \noi \hrule \vskip .2cm
 \vskip .2cm
 \leftline{\small ~~~aschieri@theorie.physik.uni-muenchen.de}
\leftline{\small ~~~castellani@to.infn.it} \leftline{\small
~~~isaevap@thsun1.jinr.ru}

\end{titlepage}
\newpage
\setcounter{page}{1}


\sect{Introduction}

The differential geometry of finite groups $G$ has proved to be a
useful tool in constructing gauge and gravity theories on discrete
spaces \cite{DMGcalculus,gravfg,tmr,CasPag,majidetal}. These
spaces are the finite group ``manifolds" associated to every
differential calculus (DC) on $G$, in 1-1 correspondence with
unions of its conjugation classes \cite{DMGcalculus,FI1,gravfg}.
The construction of DC on finite $G$ is a particular instance of a
general procedure yielding DC on Hopf algebras, first studied by
Woronowicz \cite{wor} in the noncommutative context of quantum
groups (see also \cite{ACintro} for a review with applications to
field theory).

In this paper we extend the results of \cite{gravfg,tmr,CasPag}
and formulate discretized gauge theories by means of a geometric
action $\int F \we *F$ on a finite group $G$. Using the same
geometrical tools, we also construct a discretized version of
Born-Infeld theory.

The continuum Born-Infeld theory \cite{BI}, in its commutative and
noncommutative settings, related by the Seiberg-Witten map, has
become relevant in the description of $D$-brane dynamics, see for
ex. \cite{SW}. Noncommutative structures (\cite{Connes} for a
review) in string/brane theory have emerged in the last years (see
for ex. \cite{CDS,SW} and ref.s therein), and are the object of
intense research , see e.g. \cite{dougnek} and included ref.s.
This motivates in part our investigation concerning a particularly
simple form of noncommutative Born-Infeld (BI) theory: the one
that arises by considering the BI action living on finite group
spaces.

 Here the noncommutativity is mild, in the sense that
fields commute between themselves (in the classical theory), and
only the commutations between fields and differentials, and of
differentials between themselves are nontrivial. In this framework
we obtain a discretized Yang-Mills and BI theory for every
differential calculus on a finite group.

We study the issue of duality invariance of a nonlinear
electromagnetic theory described by a generic action $S[F]$ on a
4-dimensional abelian finite group. A consistency condition is
found, and generalizes the results of the continuum limit.

Considering then Yang-Mills and BI theories as living on the
product space (D-Minkowski) $\times$ (finite group space) we
obtain, after use of Kaluza-Klein reduction techniques,
D-dimensional Yang-Mills and BI theories coupled to scalar fields.
The harmonic analysis on the discrete internal spaces is trivial.
It is tempting to interpret the product space (D-Minkowski)
$\times$ (finite group space) as a bundle of $n$ (D-1)-dimensional
branes evolving in time, $n$ being the dimension of the finite
group.

The paper is organized as follows. In Section 2, after a short
review of the differential geometry of finite groups, we show that
the metric is essentially unique (for each real differential
calculus) and then define the Hodge dual. Section 3 recalls
previously obtained results on gauge theories living on finite
group manifolds, with some new observations. In particular the
Hodge dual is used to formulate discretized gauge theories on
finite $G$ manifolds in purely geometrical terms, as in the
continuum case. In Section 4 we address the question of duality
invariance. In Section 5 we consider Yang-Mills theory on $M^D
\times G$, and reduce it via Kaluza-Klein techniques to a
continuum gauge theory in $M^D$ coupled to scalar fields. We then
specialize our analysis in Section 6 to the case $G=Z_N$. As in
the case $M^D \times Z_2$ (see for ex. \cite{tmr}), a potential
for the scalar field emerges in the $D$ - dimensional action.  In
Section 7 a finite group lattice action for Born-Infeld theory is
presented. Section 8 deals with Kaluza-Klein BI theory on $M^D
\times G$, with some explicit results for the case $M^4 \times
Z_N$. In the Appendix we discuss in more detail the Hopf algebraic
structure of finite groups $G$ and of $Fun(G)$.

\sect{A brief review of differential calculus on finite groups}

 Let $G$ be a finite group of order $n$ with
generic element $g$ and unit $e$. Consider $Fun(G)$, the set of
complex functions on $G$. An element $f$ of $Fun(G)$ is specified
by its values $f_g \equiv f(g)$ on the group elements $g$, and can
be written as
 \eq
  f=\sum_{g \in G} f_g x^g,~~~f_g \in \Cb \label{fonG}
 \en
where the functions $x^g$ are defined by
 \eq
  x^g(g') = \de^g_{g'} \label{xg}
\en
Thus $Fun(G)$ is a n-dimensional vector space, and the $n$
functions $x^g$ provide a basis. $Fun(G)$ is also a commutative
algebra, with the usual pointwise sum and product, and unit $I$
defined by $I(g)=1, \forall g \in G$. In particular: \eq x^g
x^{g'}=\de_{g,g'} x^g,~~~\sumong x^g = I \label{mul}
\en
The left and right actions of the group $G$ on itself
 \eq
\lb{dg2a} L_g \, g' = g \, g' = R_{g'} \, g \;\;\; \forall g,g'
\in G \; ,
 \en
induce the left and right actions (pullbacks) ${\cal L}_g$, ${\cal
R}_g$ on $Fun(G)$
 \eq
\lb{dg2b} [{\cal L}_g \, f ] (g') =f( g \, g') = [{\cal R}_{g'} \,
f] (g) \;\;\; \forall f \in Fun(G)\; .
 \en
 For the basis functions we find easily: \eq \Lcal_{g_1} x^{g} =
x^{g_1^{-1} g}, ~~\Rcal_{g_1} x^{g} = x^{g g_1^{-1}}
\en
Moreover:
 \eq \Lcal_{g_1} \Lcal_{g_2}=\Lcal_{g_2g_1},
~~\Rcal_{g_1} \Rcal_{g_2}=\Rcal_{g_1g_2},~~\Lcal_{g_1}
\Rcal_{g_2}=\Rcal_{g_2} \Lcal_{g_1}
 \en
 The $G$ group structure induces a Hopf algebra structure on
$Fun(G)$, and this allows the construction of differential calculi
on $Fun(G)$, according to the techniques of ref.
\cite{wor,ACintro}. We summarize here the main properties of these
calculi. A detailed treatment can be found in \cite{gravfg}, and
Hopf algebraic formulas, allowing contact with the general method
of \cite{wor,ACintro}, are listed in the Appendix.
 \sk
 A (first-order) differential
calculus on $Fun(G)$ is defined by a linear map $d$: $Fun(G)
\rightarrow \Gamma$, satisfying the Leibniz rule $
 d(ab)=(da)b+a(db),~~\forall a,b\in Fun(G)$.
The ``space of 1-forms" $\Ga$ is an appropriate bimodule on
$Fun(G)$, which essentially means that its elements can be
multiplied on the left and on the right by elements of $Fun(G)$.
{} From the Leibniz rule $da=d(Ia)=(dI)a+Ida$ we deduce $dI=0$.
Consider the differentials of the basis functions $x^g$. From
$0=dI=d(\sumong x^g)=\sumong dx^g$ we see that in this calculus
only $n-1$ differentials are independent.
 \sk

A differential calculus is
 {\sl left or right covariant} if the left or right action of
 $G$ ($\Lcal_g$ or $\Rcal_g$) commutes with the exterior derivative $d$.
 Requiring left and right covariance in fact {\sl defines} the action of
 $\Lcal_g$ and $\Rcal_g$ on differentials: $\Lcal_g db \equiv
 d(\Lcal_g b), \forall b \in Fun(G)$ and similarly for
 $\Rcal_g db$. More generally, on elements of $\Ga$
 (one-forms) we define $\Lcal_g$ as:
 \eq
 \Lcal_g (adb) \equiv (\Lcal_g a) \Lcal_g db =
 (\Lcal_g a) d (\Lcal_g b)
 \en
 and similarly for $\Rcal_g$.
 A differential calculus is called {\sl bicovariant} if it is
both left and right covariant.
 \sk
 As in usual Lie group manifolds, we can introduce in $\Ga$
 the left-invariant one-forms $\theta^g$:
 \eq
  \theta^g \equiv
\sumonh x^{hg^{-1}} dx^h =\sumonh x^h dx^{hg}, \label{deftheta}
\en
These $\theta^g$ correspond to the $\theta^{g^{-1}}$ of ref.s
\cite{DMGcalculus,gravfg,tmr,CasPag}.

 It is immediate to check
that indeed $\Lcal_k \theta^g = \theta^g$.
 The right action of $G$ on the elements
$\theta^g$ is given by:
 \eq
  \Rcal_h \theta^g =
\theta^{ad(h)g},~~\forall h \in G \label{Rontheta}
\en
where $ad$ is the adjoint action of $G$ on itself, i.e. $ad(h)g
\equiv hgh^{-1}$. Notice that $\theta^e$ is biinvariant, i.e. both
left and right invariant.

 {} From $\sumong dx^g=0$ one finds: \eq
 \sumong
\theta^g = \sum_{g,h \in G} x^h dx^{hg}= \sumonh x^h \sumong
dx^{hg}=0 \label{sumtheta}
\en
Therefore we can take as basis of the cotangent space $\Ga$ the
$n-1$ linearly independent left-invariant one-forms $\theta^g$
with $g \not= e$ (but smaller sets of $\theta^g$ can be
consistently chosen as basis, see later). Using (\ref{mul}) the
relations (\ref{deftheta}) can be inverted:
 \eq
  dx^h = \sumong x^{hg^{-1}}\theta^g =
   \sumongnote (x^{hg^{-1}} - x^h)\theta^g \label{dxastheta}
\en
 Analogous results hold
for right invariant one-forms $\zeta^g$:
 \eq
  \zeta^g = \sumonh x^{g^{-1}h}dx^h
\en
Using the definition of $\theta^g$ (\ref{deftheta}), the
commutations between $x$ and $\theta$ are easily obtained:
 \eq
  x^h dx^g = x^h \theta^{h^{-1}g} =
\theta^{h^{-1}g} x^g ~~(h\not=g)~~\Rightarrow \theta^g x^h=
x^{hg^{-1}}\theta^g ~~(g\not=e) \label{xthetacomm}
\en
and imply the general commutation rule between functions and
left-invariant one-forms:
 \eq
   \theta^g f = [\Rcal_g f] \theta^g~~~~~~~(g \not= e)
    \label{fthetacomm}
 \en
Thus functions do commute between themselves (i.e. $Fun(G)$ is a
commutative algebra) but do not commute with the basis of
one-forms $\theta^g$. In this sense the differential geometry of
$Fun(G)$ is noncommutative.
 \sk
The differential of an arbitrary function $f \in Fun(G)$ can be
found with the help of (\ref{dxastheta}):
 \eqa
& & d f = \sum_h \, f_h \, d  x^h = \sum_{g,h} f_h \, x^{h \,
g^{-1}} \, \theta^g  = \sum_{g \neq e} \,  ( \sum_h f_h \, x^{h \,
g^{-1}} - f ) \, \theta^g = \nonumber \\ & &~~~~ = \sum_{g \neq e}
\, ( [{\cal R}_g \, f] - f )  \, \theta^g = \sum_{g \neq e} \,  (
t_g \, f) \, \theta^g \; . \label{partflat}
 \ena
\noi Here the finite difference operators  $t_g = \Rcal_g - 1$ are
the analogues of (left-invariant) tangent vectors. They satisfy
the Leibniz rule:
    \eq
 t_g (ff')=
 (t_g f)  f'  +\Rcal_g (f) t_g f'
 =(t_g f) \Rcal_{g} f'  + f t_g f' \label{tgLeibniz}
 \en
 \noi and close on the fusion algebra:
 \eq
 t_g \, t_{g'} = ({\cal R}_{gg'}-1) - ({\cal R}_{g}-1) - ({\cal
R}_{g'}-1) = \sum_h \, C^h_{g,g'} \, t_h \; ,
 \label{fusion}
 \en
  \noi  where the
structure constants $C^h_{g, g'}$ are
 \eq
 C^h_{g,g'} = \delta^h_{g g'} - \delta^h_g - \delta^h_{g'} \; ,
  \label{cconst}
   \en
The commutation rule  (\ref{fthetacomm}) allows to express the
differential of a
 function $f \in Fun(G)$ as a commutator of $f$ with the biinvariant form $\sum_{g \neq
e} \theta^g= -\theta^e $:
 \eq d f = [ \sum_{g \neq e} \theta^g , \, f]= -[\theta^e,f] \; .
 \label{df}
 \en
 An {\sl exterior product}, compatible with the left and right
actions of $G$, can be defined as
 \eqa & &
 \theta^g \wedge \theta^{g'}
 = \theta^g \otimes \theta^{g'} - \sum_{k,k'} \Lambda^{g \, g'}_{~~k'
\, k}  \theta^{k'} \otimes \theta^k = \theta^g \otimes \theta^{g'}
-  \theta^{g g' g^{-1}} \otimes \theta^g = \nonumber\\
 & & ~~~~~~ = \theta^g
\otimes \theta^{g'} - [{\cal R}_g \theta^{g'} ] \otimes \theta^g
\; , \;\;\; (g,g' \neq e) \; , \label{extheta}
 \ena
 where the tensor product between elements $\rho,\rhop \in \Ga$ is
defined to have the properties $\rho a\otimes \rhop=\rho \otimes a
\rhop$, $a(\rho \otimes \rhop)=(a\rho) \otimes \rhop$ and $(\rho
\otimes \rhop)a=\rho \otimes (\rhop a)$. The braiding matrix
$\La$:
 \eq
\La^{g \, g'}_{~~k' \, k} = \delta^{g g' g^{-1}}_{k'} \,
\delta^g_k \; , \;\;\; (g,g' \neq e) \; .
 \en
 satisfies the Yang-Baxter equation $\L{nm}{ij} \L{ik}{rp} \L{js}{kq}
 =\L{nk}{ri} \L{ms}{kj}\L{ij}{pq}$. With this exterior product we find
 \eq
\theta^g \wedge \theta^g = 0 \;\;\; (\forall g) \; , \;\;\;
\theta^g \wedge \theta^{g'} = - \theta^{g'} \wedge \theta^{g}
\;\;\; (\forall g,g': \;\; [g, \, g']=0 \; , \;\; g \neq e) \; .
 \en
 Moreover
 \eq
 \theta^g \wedge \theta^{g'} ~f = (\Rcal_{gg'} f) \theta^g \wedge \theta^{g'}
 \label{commf}
 \en
 Left and right actions on $\Ga \otimes \Ga$ are
  simply defined by:
  \eq
  \Lcal_h (\rho \otimes \rhop)= \Lcal_h \rho \otimes \Lcal_h
  \rhop,~~~
\Rcal_h (\rho \otimes \rhop)= \Rcal_h \rho \otimes \Rcal_h
  \rhop
  \en
 Compatibility  of the exterior product with $\Lcal$ and $\Rcal$
 means that
 \eq
 \Lcal(\theta^i \we \theta^j)=\Lcal\theta^i \we \Lcal
 \theta^j, ~~\Rcal(\theta^i \we \theta^j)=\Rcal\theta^i \we \Rcal
 \theta^j
 \en
 Only the second relation is nontrivial and is verified upon use
 of the definition (\ref{extheta}).
 We can generalize this definition
 to exterior products
 of left-invariant $1$-forms as follows \cite{AB}:
\eq
\theta^{i_1} \we ... \we \theta^{i_k} \equiv
A^{i_1..i_k}_{j_1..j_k}~
 \theta^{j_1} \otimes ...\otimes
\theta^{j_k} \label{multiwedge}
\en
\noi or in short-hand notation (FRT matrix notations \cite{FRT}):
\eq \theta^{1} \we ... \we \theta^{k}= A_{1...k}~
 \theta^{1} \otimes ...\otimes
\theta^{k}\label{not1}
\en
\noi The labels $1...k$ in $A$ refer to index couples, and
$A_{1,...k}$ is the analogue of the antisymmetrizer of $k$ spaces,
defined by the recursion relation
 \eq
  A_{1\ldots k} = (1 - \Lambda_{k-1,k} +
\Lambda_{k-2,k-1} \Lambda_{k-1,k}-   \ldots -(-1)^k \Lambda_{12}
\Lambda_{23} \cdots \Lambda_{k-1,k}) A_{1\ldots k-1} ,
\label{Asymm}\label{not2}
\en
 where $A_{12} = 1-\La_{12}$.
  The space of $k$-forms $\Ga^{\we k}$ is therefore
defined as in the usual case but with the new permutation operator
$\La$, and can be shown to be a bicovariant bimodule (see for ex.
\cite{Athesis}), with left and right action defined as for $\Ga
\otimes ...\otimes \Ga$ with the tensor product replaced by the
wedge product. The property (\ref{commf}) generalizes to:
 \eq
 \theta^{i_1} \we ... \we \theta^{i_k} ~f = (\Rcal_{i_1...i_k} f)
  \theta^{i_1} \we ... \we \theta^{i_k}
  \en
 The graded bimodule $\Omega = \sum_k \Ga^{\we k}$ is the exterior
algebra of forms. As recalled in the Appendix, this algebra is
also a Hopf algebra \cite{Brz,IsO}.

The {\sl exterior derivative} is defined as a linear map $
d~:~\Gamma^{\we k} \rightarrow \Gamma^{\we (k+1)}$ satisfying
$d^2=0$ and the graded Leibniz rule
 \eq
 d(\rho \we \rhop)=d\rho \we \rhop +
(-1)^k \rho \we d\rhop \label{propd1}
\en
\noi where $\rho \in \Ga^{\we k}$, $\rhop \in \Ga^{\we k'}$,
$\Ga^{\we 0} \equiv Fun(G)$ . Left and right action is defined as
usual:
 \eq
 \Lcal_g (d\rho)=d \Lcal_g \rho, ~~~\Rcal_g (d\rho)=d \Rcal_g \rho
 \label{propd2}
\en
In view of relation (\ref{Rontheta}), and (\ref{Hopfomega}), the
algebra $\Omega$ has natural quotients over the ideals $H_g$= $\{
\theta^{hgh^{-1}}, \forall h \}$, corresponding to the various
conjugacy classes of the elements $g$ in $G$. The different
bicovariant calculi on $Fun(G)$ are in 1-1 correspondence with
different quotients of $\Omega$ by any sum of the ideals $H=\sum
H_g$, cf. \cite{DMGcalculus,FI1,gravfg}. In practice one simply
sets $\theta^g = 0$ for all $g \not= e$ not belonging to the
particular union $G'$ of conjugacy classes characterizing the
differential calculus. The dimension of the space of independent
1-forms for each bicovariant calculus on $Fun(G)$ is therefore
equal to the dimension of the subspace $\Ga/H$.
 \sk
 The {\sl Cartan-Maurer equation} for the differential forms $\theta^g$
(\ref{deftheta}) is obtained by direct calculation, using the
definition (\ref{deftheta}), the expression  (\ref{dxastheta}) of
$dx^h$ in terms of $\theta$'s, and the commutations
(\ref{xthetacomm}):
 \eq
   d \theta^g = - \sum_{h \not= e,h'\not= e} \de^g_{hh'} \theta^h \we \theta^{h'} +
 \sum_{k \not= e} \theta^k \we \theta^g + \sum_{k \not= e} \theta^g \we \theta^k =
- \sum_{h \neq e} \sum_{h' \neq e} \, C^g_{h,h'} \,  \theta^h
\wedge \theta^{h'} \; ,~~(g\not=e) \label{CM}
 \en
  where the structure constants $C^g_{h,h'}$ are given in (\ref{cconst}).
Using  the identity:
 \eq
  \sum_{h \not= e,h'\not= e} \, \delta^k_{hh'}
\, \theta^h \wedge \theta^{h'} = \sum_{h \not= e,h'\not= e} \,
\delta^k_{hh'} \, \left( \theta^h \otimes \theta^{h'}-
 \theta^{h h'h^{-1}} \otimes \theta^h \right) = 0 \label{iden}
 \en
 \noi the Cartan-Maurer equation can be rewritten by means of the
anticommutator of $\theta^g$ with the biinvariant form $\theta^e$:
 \eq
 d\theta^g=-\theta^e \we \theta^g - \theta^g \we \theta^e
 \label{dtheta}
 \en
 cf. the case of 0-forms (\ref{df}). Considering now a generic
 element $\rho = a \theta$ of $\Gamma$ it is easy to find that
 $d\rho = -\theta^e \we \rho - \rho \we \theta^e$. The general
 rule is
 \eq
 d\rho = [-\theta^e, \rho]_{grad} \equiv
 -\theta^e \we \rho + (-1)^{deg(\rho)} \rho \we \theta^e
 \en
 valid for any $k$-form, where $[-\theta^e, \rho]_{grad}$ is the
 graded commutator. Thus $-\theta^e = \sum_{k \not= e} \theta^k$
 is the $X$ generator of Woronowicz theory \cite{wor}, or BRST operator
 \cite{IsO}; in our case $X$ belongs to $\Gamma$.
 \sk
 There are two (Hopf
algebra) {\sl conjugations} on $Fun(G)$ \cite{DMGcalculus,CasPag}
 \eq
(x^g)^* = x^g \; , \;\;\; (x^g)^\star = x^{g^{-1}}
 \label{conj}
 \en
On one-forms
 \eq
  (\theta^g)^*= - \theta^{g^{-1}} \; ,
  \;\;\; (\theta^g)^\star = \zeta^{g} \;,
   \en
with $(f\th^g h)^*=h^*(\th^g)^* f^*$, $(f\th^g h)^\star=h^\star(\th^g)^\star
f^\star$.
These involutions can be extended to the whole tensor algebra of one-forms
(\cite{wor}):
\eq
(\theta^{i_1} \otimes \theta^{i_2} \otimes \cdots \otimes \theta^{i_k})^*=
\Pi^{i_1,\ldots i_k}_{j_1,\ldots j_k}
(\theta^{j_k})^* \otimes (\theta^{j_{k-1}})^* \otimes \cdots \otimes (\theta^{j_1})^*~~,
\en
here the linear operator $\Pi^{i_1,\ldots i_k}_{j_1,\ldots j_k}$ is the permutation
$\left({}^{{}{1,\:2\,,\,\ldots\, k-1,k}}_{{}{k,k-1,\ldots\, 2\,,\,1}}\right)$ where nearest neighbour
transpositions are replaced with the braid operator $\Lambda^{-1}$. A realization of
$\Pi_{1,\ldots k}$ is
$\Pi_{1,\ldots k}=\Lambda_{1,2}^{-1}\,~\Lambda_{2,3}^{-1}\Lambda_{1,2}^{-1}\,~
\Lambda_{3,4}^{-1}\Lambda_{2,3}^{-1}\Lambda_{1,2}^{-1}\,~\ldots\,~
\Lambda_{k-1,k}^{-1}\Lambda_{k-2,k-1}^{-1}{\!}\ldots\Lambda_{2,3}^{-1}\Lambda_{1,2}^{-1}
\,$ [see (\ref{not1}),(\ref{not2}) for the notation]. Explicitly:
\eq
(\theta^{i_1} \otimes \theta^{i_2} \otimes \cdots \otimes
 \theta^{i_k})^*= (-1)^k~
 \theta^{ad(i_2...i_k)^{-1} i^{-1}_1} \otimes
 \theta^{ad(i_3...i_k)^{-1} i_2^{-1}} \otimes \cdots\otimes
 \theta^{i_k^{-1}}
\label{startensor}
\en
In particular, this antilinear involution maps exterior forms onto exterior forms,
and we have
\eq
 (\theta^{i_1} \we \theta^{i_2} \we \cdots \we
 \theta^{i_k})^*
= (-1)^{k(k+1)\over 2}~\theta^{i_k^{-1}}\we \cdots \we \theta^{i_2^{-1}}
 \we \theta^{i_1^{-1}} \label{starwedge}
  \en
and more generally $(\rho \we \rhop)^* = (-1)^{deg(\rho) deg(\rhop)}
\rhop^* \we \rho^*$ with $\rho,\,\rhop$ arbitrary forms.
These same formulas hold also when we consider the ${}^\star$-conjugation.
We'll use the *-conjugation in the sequel. Consistency of this conjugation
requires that if $\theta^g \neq 0$ then $\theta^{g^{-1}} \neq 0$ as well:
we have to include in $\Ga/H$ at least the two
ideals $H_g$ and $H_{g^{-1}}$ (if they do not coincide). We obtain
thus a ${}^*$-differential calculus, i.e. $(df)^*=d ( f^*)$.
 \sk
  The fact that both $\theta^g$ and $\theta^{g^{-1}}$ are included in
  the basis of left-invariant 1-forms characterizing
   the differential calculus also ensures the existence of a unique metric
    (up to a normalization).

  The {\sl metric} is defined as a bimodule pairing, symmetric
  on left-invariant 1-forms. It maps couples of 1-forms
  $\rho,\sigma$ into $Fun(G)$, and satisfies the properties
 \eq
  <f\rho,\sigma h>=f<\rho,\sigma>h~,~~<\rho f,\sigma>
  =<\rho, f\sigma>~. \label{metricproperties}
 \en
 where $f$ and $h$ are arbitrary functions belonging to $Fun(G)$.
 Up to a normalization the above properties
 determine the metric on the left-invariant 1-forms. Indeed from
 $<\theta^g,f\theta^h>=<\theta^g,\theta^h>\Rcal_{h^{-1}}f= \Rcal_g
 f <\theta^g,\theta^h>$ one deduces:
 \eq
 g^{rs} \equiv <\theta^r,\theta^s> \equiv -\de^r_{s^{-1}} \label{metric}
 \en
 Thus $g^{rs}$ is symmetric and $\theta^r$ has nonzero
  pairing only with ${\theta^{r^-1}}$.
 The pairing is compatible with the ${}^*$-conjugation
  \eq
 <\rho,\sigma >^*=<\sigma^*,\rho^*> \label{starmetric}
  \en
 We can generalize  $<\!\!~,\!\!~>$  to
 tensor products of $k$ left-invariant one-forms:
\eq
 < \theta^{i_1} \otimes \cdots \otimes \theta^{i_k},
  \theta^{j_1} \otimes \cdots \otimes \theta^{j_k}> \equiv
< \theta^{i_1}, \theta^{j'_1}> < \theta^{i_2}, \theta^{j'_2}>\ldots
 < \theta^{i_k}, \theta^{j'_k}>  \label{pairingtensors}~
\en
where $j'_\ell=(i_{\ell+1}i_{\ell+2}...i_k) j_\ell
(i_{\ell+1}i_{\ell+2}...i_k)^{-1}$, i.e. $\th^{j'_\ell}={\cal
R}_{i_{\ell+1}i_{\ell+2}...i_k}\th^{j_\ell}$, and $j'_k = j_k$.
   Using (\ref{startensor}) the above definition is equivalent to
  \eq
   < (\theta^{i_1} \otimes \cdots \otimes \theta^{i_k})^*,
  \theta^{j_1} \otimes \cdots \otimes \theta^{j_k}>=\delta^{i_1j_1}\ldots
\delta^{i_kj_k}
  \label{dualitypairing}
  \en
{}From this formula we see that the pairing is symmetric on left-invariant tensors
(tensor products of left-invariant one-forms).
  The pairing (\ref{pairingtensors}) is extended to all tensors
  by $ <f\rho,\sigma h>=f<\rho,\sigma>h$ where now $\rho$ and
  $\sigma$ are generic tensors of same order. Then we prove easily
  that $<\rho f,\sigma> =<\rho, f\sigma>$ for any function $f$
  so that $<\!\!~,\!\!~>$ is a bimodule pairing. Left and right
  invariance of $<~,~>$ is straightforward. Moreover the pairing is compatible
  with the ${}^*$-conjugation: formula (\ref{starmetric}) holds also
when $\rho$ and $\sigma$ are arbitrary tensors. \sk

In general for a differential calculus with $m$ independent
tangent vectors, there is an integer $p  \geq m$ such that the
linear space of left-invariant $p$-forms is 1-dimensional, and $(p+1)$- forms
vanish identically \footnote{with the exception of $Z_2$, see ref.
\cite{tmr}}. This means that every product of $p$ basis one-forms
$\theta^{g_1} \we \theta^{g_2} \we ... \we \theta^{g_p}$ is
proportional to one of these products, that can be chosen to
define the volume form $vol$:
 \eq
 \theta^{g_1} \we \theta^{g_2} \we ... \we \theta^{g_p}=
 \epsilon^{g_1,g_2,...g_p}\: vol \label{vol}
 \en
 where $\epsilon^{g_1,g_2,...g_p}$ is the proportionality
 constant, a real number since the braiding matrix $\La$ is real.

 The volume $p$-form is obviously left invariant. It is also right invariant
 \cite{gravfg} (the proof is based on the $ad(G)$ invariance of
 the $\epsilon$ tensor:
 $\epsilon^{ad(g)h_1,...ad(g)h_p}=\epsilon^{h_1,...h_p}$).
 \sk
 Finally, if $vol = \theta^{k_1} \we ...\we \theta^{k_p}$, then
 \eq
 vol^*=(-1)^{p (p+1) \over 2}\epsilon^{k_p^{-1}...k_1^{-1}} vol
 \en
 so that $vol$ is either real or imaginary. If $vol^* = -vol$  we can always
 multiply it by $i$ and obtain a real volume form. In that case
 comparing $(\theta^{g_1} \we ...\we \theta^{g_p})^*
 =(-1)^{p(p+1)\over 2} \theta^{g_p^{-1}}\we ... \we \theta^{g_1^{-1}} =
 \epsilon^{g_p^{-1}...g_1^{-1}} (-1)^{p(p+1)\over 2} vol$
 with $(\theta^{g_1} \we ...\we \theta^{g_p})^* =
 \epsilon^{g_1...g_p} ~(vol)^* =  \epsilon^{g_1...g_p} ~vol$
 yields
 \eq
 \epsilon^{g_p^{-1}...g_1^{-1}} = (-1)^{p(p+1)\over 2} ~\epsilon^{g_1...g_p}
 \label{epsiminus}
 \en

 Computing the pairing of $vol$ with itself yields:
 \eq
 <vol,vol>={\cal N},~~~~~
  {\cal N}\in  \mathbb{R}
   \label{volvol}
 \en
 Clearly the pairing (\ref{pairingtensors}) or $vol$ can be
 normalized so that $<vol,vol>= \pm 1$ but we'll use (\ref{volvol}) in
 the following.
 \sk
Having identified the volume $p$-form it is natural to define the
integral of the generic $p$-form $f~vol$ on $G$
 \eq
 \int f ~vol  = \sumong f(g) \label{intpform}
 \en
 the right-hand side being just the Haar measure of the function $f$.
\sk
 The {\sl Hodge dual}, an important ingredient for gauge theories,
 can be defined as the unique map from $k$-forms $\sigma$ to $(p-k)$-forms
 $* \sigma$ such that (see \cite{AB} for a similar construction):
\begin{equation}
\label{defhodge} \rho\wedge *\sigma = <\rho,\sigma> vol
~~~~~~~~~\rho,\sigma  {\mbox{ $n$-forms}}
\end{equation}
The Hodge dual is left linear; if $vol$ is central it is also
right linear :
 \eq *(f\rho\,h)=f(*\rho)h
\label{linearityhodge}
 \en
 with $f,h \in Fun(G)$.
 Moreover
 \eq
 {}*\Ncal= \Ncal~vol~~,~~*vol=\Ncal~~.
\label{compatibilehodge}
 \en
 \noi {\bf Note:}
 the ``group manifold" of a finite group is simply a
  collection of points corresponding to the
group elements, linked together in various ways, each
corresponding to a particular differential calculus on $Fun(G)$
\cite{DMGcalculus,gravfg}. The links are associated to the tangent
vectors $\Rcal_h -1$ of the differential calculus, or equivalently
to the right actions $\Rcal_h$, where $h$ belongs to the union
$G'$ of conjugacy classes characterizing the differential
calculus. Two points $x^g$ and $x^{g'}$ are linked if
$x^{g'}=\Rcal_h x^g$, i.e. if $g'=gh^{-1}$  for some $h$ in $G'$.
The link is oriented from $x^g$ to $x^{g'}$ (unless $h=h^{-1}$ in
which case the link is unoriented): the resulting ``manifold" is
an oriented graph. From every point exactly $m$ (= number of
independent 1-forms) links originate. Some examples of finite
group manifolds can be found in \cite{gravfg}.


\sect{Gauge theories on finite groups}


A natural question arises: is it possible to construct field
theories on the discrete spaces provided by finite group manifolds
? The answer is affirmative: exploiting the differential calculus
on finite $G$, gauge and gravity theories have been constructed in
ref.s \cite{DMGcalculus,gravfg,tmr,CasPag,majidetal}. To prepare
the ground for discretized Born-Infeld theory, we recall in this
section how to construct $\Gcal = U(N)$ gauge field theory on
finite $G$ group spaces and add some new observations. The
discretized Yang-Mills action involves only geometric objects: the
2-form field strength $F$ , the $*$-Hodge operator and the
invariant integral on $G$.
 \sk
  The gauge field of a Yang-Mills
theory on a finite group $G$ is a matrix-valued one-form
$A(x)=A_h(x) \theta^h$. The components $A_h$ are functions on $G$:
they can be considered functions of the ``coordinates" $x^g$,
since any element of $Fun(G)$ can be expanded on the basis
functions $x^g$. Moreover they are matrix-valued, i.e.
$A=(A_h)^\al_{~\be} \theta^h$, $\al ,\be=1,...N$. In the following
matrix multiplication is implicit.

As in the usual case, $\Gcal$ gauge transformations are defined as
  \eq
  A'=-(dT)T^{-1}+TAT^{-1} \label{gaugeA}
  \en
  where $T(x)=T(x)^{\al}_{~\be}$ is an $N\times N$ representation of a $\Gcal$
  group element; its matrix entries belong to $Fun(G)$.
    In components:
  \eq
  A'_h=-(t_h T) \Rcal_{h}
  T^{-1}+TA_h \Rcal_{h}T^{-1}=
  T \, [ 1 + A_h ] \,  [{\cal R}_h T^{-1}] - 1
  \label{gaugeAcomp}
  \en
  Matter fields $\psi_\al(x)$ transform in a representation of $\Gcal$ as
  $\psi\rightarrow \psi'=T\psi$,
   and their covariant derivative, defined by
  \eq
  D\psi= d\psi +A \psi,
  \en
  transforms homogeneously: $(D\psi)'=T (D\psi)$.
The 2-form field strength $F$ arises as usual in the square of the
covariant derivative $(D^2) \psi=F \psi$; it is given by the
familiar expression $F=dA+A\we A$, which in components takes the
form:
 \eqa
  F&=&dA+A\we A=d(A_k\theta^k)+A_h\theta^h \we A_k
 \theta^k\nonumber \\
  &=&(t_h A_k)~ \theta^h\we\theta^k + A_k d\theta^k + A_h
 (\Rcal_{h} A_k) ~\theta^h \we \theta^k  \nonumber\\
  &=& [t_h A_k - A_j C^j_{~hk} + A_h (\Rcal_{h} A_k)]~
 \theta^h \we \theta^k \label{F}
 \ena
 and satisfies the Bianchi identity:
 \eq
 d \, F + A \we F - F \we A =  0 \label{FBianchi}
 \en
 Note that $A \we A \not= 0$ even if the gauge group $\Gcal$ is
 abelian. Thus $U(1)$ gauge theory on a finite group space
 looks like a nonabelian theory,
 a situation occurring also in noncommutative $*$-deformed gauge
 theories.

  Under gauge transformations (\ref{gaugeA}) $F$ varies homogeneously:
 \eq
 F'=TFT^{-1}
\label{Fgaugetr}
 \en

The gauge variation (\ref{gaugeAcomp}) suggests the definition of
the link fields $U_h(g)$ and link 1-form $U(g)$:
 \eq
 U_h = 1 + A_h \; , \;\;\;
U = \sum_{h \neq e} U_h \, \theta^h = \sum_{h \neq e} \, \theta^h
+ A \label{defU}
\en
transforming homogeneously:
 \eq U'_g = T \, U_g \, [{\cal R}_g
T^{-1}] \; , \;\;\; U' = T \, U \, T^{-1} \; .
\en
\noi The field strength (\ref{F}) can be expressed in terms of the
link fields:
 \eq
 F= \left( U_h \, [ {\cal R}_h U_k]  -
 U_{g} \delta^g_{hk}  \right) \, \theta^h \wedge \theta^k=
   U_h \,  ({\cal R}_h U_k) \, \theta^h \wedge \theta^k = U \we
  U \label{FasU2}
 \en
 where we have used the identity (\ref{iden}).

 Defining the components $F_{h,g}$ as:
 \eq
  F\equiv F_{h,k}~ \theta^h \otimes \theta^k\label{Fcomp}
  \en
  eq. (\ref{FasU2}) yields:
  \eq
  F_{h,k} = U_h \, ({\cal R}_h \, U_k) -
  U_k ({\cal R}_k \, U_{k^{-1}hk}) \; . \label{FasUcomp}
  \en
\sk The Yang-Mills action is the geometrical action quadratic in
$F$ given by
 \eq
 A_{YM}=-\int Tr (F \we *F)=-\int Tr <F,F> vol
 \label{YMact}~\;.
 \en
 Recalling
 the pairing properties (\ref{metricproperties}) the proof
 of gauge invariance of
 $ Tr <F,F>$ is immediate:

 $Tr <F',F'>=Tr<TFT^{-1},TFT^{-1}>=Tr T<F,F>T^{-1}=Tr<F,F>~.$
\sk The metric (\ref{metric}) is an euclidean metric (as is easily
seen using a real basis of one-forms) and as usual we require
(\ref{YMact}) to be real and positive definite. This restricts the
gauge group $\Gcal$ and imposes reality conditions on the gauge
potential $A$. Since (cf. (\ref{intpform})) $\int Tr <F,F>
vol=\sum_G Tr <F,F>$, positivity of (\ref{YMact}) requires $ -(Tr
<F,F>) \geq 0$. Explicitly
 \eqa & &
  <F,F> = <F_{r,s} \theta^r \otimes \theta^s, F_{m,n}
 \theta^m \otimes \theta^n>  = F_{r,s} R_{rs}F_{m,n}<\theta^r \otimes \theta^s,
\theta^m \otimes \theta^n> \nonumber \\
&~ & ~~~~~~~~~~~= F_{r,s} R_{rs}F_{m,n} \delta^{s^{-1}}_n\delta^{r^{-1}}_{s\,m\,s^{-1}}
=  F_{r,s}
\Rcal_{rs} F_{s^{-1}r^{-1}s,s^{-1}}
 \ena
and \eq -Tr   <F,F>= -(F_{r,s})^\al_{~\be} (\Rcal_{rs}
F_{s^{-1}r^{-1}s,s^{-1}})^\be_{~\al}~.\label{tracciaFF}
\en
We see that (\ref{tracciaFF}) is positive definite if
$\,-(\Rcal_{rs}
F_{s^{-1},r^{-1}})^\be_{~\al}=(F_{r,s}^*)^\al_{~\be}$, i.e. if
 \eq
 F_{r,s}^\dagger=  -\Rcal_{rs} F_{s^{-1}r^{-1}s,s^{-1}}
 ~~\label{Freal}
 \en
The $\dagger$
conjugation by definition acts as hermitian
 conjugation on the matrix structure, and as the ${}^*$-conjugation (introduced
 in the previous Section) on the $Fun(G)$ entries of the matrix.
%
%
 The gauge action then can be rewritten (sum on the indices
$h,k,\al,\be$ understood)
 \eq
 A_{YM}=\sum_G Tr (F_{h,k} F^\dagger_{h,k})=
\sum_G ({F_{h,k}})^\al_{~\be} {(F^*_{h,k})}^{\al}_{~\be}
\label{YMaction}~.
 \en
{}From (\ref{FasUcomp})  or  (\ref{F}) we see that (\ref{Freal}) holds if
\eq
  A^\dagger_h=\Rcal_{h} A_{h^{-1}}
~~\mbox{ or equivalently: }~~~
U^\dagger_h=\Rcal_{h} U_{h^{-1}}~.\label{UArealitycomp}
\en
These relations simply state that the one forms $A$ and $U$ are
antihermitian:
\eq
A^\dagger = -A~~~,~~~~~
U^\dagger = -U~~, \label{UAreality}
\en
hermitian conjugation on matrix valued one-forms
$A$ (or $U$) being defined as:
  \eq
  A^\dagger = (A_h \theta^h)^\dagger =
(\theta^h)^*
  A_h^\dagger=-\theta^{h^{-1}} A^\dagger_h=-\theta^h
  A^{\dagger}_{h^{-1}}=-({\cal R}_h A^\dagger_{h^{-1}})\,\th^h~.
  \en
{}From (\ref{UAreality}), (\ref{FasU2}) and (\ref{starwedge}) we
obtain that the 2-form $F$ is antihermitian $F=-F^\dagger$. \sk
{}Finally gauge trasformations must preserve the hermiticity
properties of $A$ and $F$ and this is the case if the
representation $T$ of $\Gcal$ is unitary. We thus conclude that
the action (\ref{YMact}) has maximal gauge group $\Gcal=U(N)$.

Notice that writing
$A=(A_h){}^\al_{\;\be}\th^h$, where $\al,\be=1,...N$, the reality
condition $A^\dagger=-A$ is equivalent to
$\overline{A_h(g)}{}^\be_{\;\al}=A_{h^{-1}}(gh)^\al_{\;\be}\,$
and is thus not local (not fiberwise).
It follows that
$A$ has values in $M_{N\times
  N}({\mbox{\bf{C}}})$ and not in
the Lie algebra of $U(N)$.
Nevertheless  $A^\dagger=-A$ is a good reality condition because it cuts
by half the total number of components of $A$. This can be seen
counting the real components of the antihermitian field $A$: they are
$N^2\times n\times d$, where $n$ is the number of
points of $G$, and the ``dimension'' $d$ counts the number of independent
left-invariant one-forms.
In conclusion, when $\Gcal=U(N)$, we have a bona fide pure gauge
action, where the number of components of $A$ is consistent with
the dimension of the gauge group.

Note also from
(\ref{gaugeAcomp}) that since $t_h T$ is a finite difference of
group elements, then $A_h$ cannot belong to
 the $\Gcal$ Lie algebra.\sk

 The action (\ref{YMaction}) can be expressed in terms of the link fields
 $U_h$: substituting (\ref{FasUcomp}) into (\ref{YMaction}) leads to
 \eq
 A_{YM} = 2 \sum_G Tr [U_k U^\dagger_k U_h U^\dagger_h-U^\dagger_k U_h
 (\Rcal_h U_k)(\Rcal_k U^\dagger_{k^{-1}hk})] \label{YMactionU}
 \en
 after use of the cyclic property of $Tr$, and of (\ref{UArealitycomp}).
 Moreover we also needed the obvious properties
 \eqa
 & &\sum_G f = \sum_G \Rcal_k f~~~~~~\forall k,~\forall f \in Fun(G)\\
 & &\sum_h f_h = \sum_h f_{k^{-1}hk} ~~~~\forall k,~\forall f_h \in
 Fun(G)
 \ena

 When the finite group $G$ is abelian, the action
 (\ref{YMactionU}) reduces to a Wilson-like action if
we require $U_h$ to be unitary:
 $U_h^{\;\dagger}=(U_h)^{-1}$.
See also ref.  \cite{DMgauge} for the case
$G= Z_N \times \cdots \times Z_N $.
 \sk
 \noi {\bf Note 1.}
Here we sketch an equivalent presentation of the Yang-Mills
action. We first define it to be the geometric action
 \eq A_{YM}=
\int Tr (F \we *\: F^\dagger)= \int Tr <F,F^\dagger> vol
 \label{YMactcompl}~\;.
\en
so that $A_{YM}$ is trivially real and invariant under gauge
transformations
(\ref{Fgaugetr}) provided that $T^\dagger=T^{-1}$.
We then impose reality conditions
on $A$, and therefore on $U$ and $F$, that are preserved under $U(N)$ gauge
transformations. We thus obtain a pure gauge theory.
\sk
 \noi {\bf Note 2.} One can construct the finite group lattice analogues
of Wilson loops. Indeed, consider the exterior product $U^k$ :
 \eqa & &
 U^k =  U_{h_1} \, \theta^{h_1} \we U_{h_2} \,
\theta^{h_2} \we \dots \we U_{h_k} \,  \theta^{h_k} = \nonumber \\
 & & ~~~~~=
U_{h_1} [{\cal R}_{h_1} U_{h_2}] \, [{\cal R}_{h_1 h_2} U_{h_3}]
\dots [{\cal R}_{h_1 \dots h_{k-1}} U_{h_k}]
 \theta^{h_1} \we  \theta^{h_2} \we \,
\dots \,  \we \theta^{h_k} \equiv \nonumber\\
 & & ~~~~~\equiv
 U_{h_1, \dots  ,h_k}
 \theta^{h_1} \we   \,
\dots \,  \we \theta^{h_k}
 \ena
 (no sum on the indices $h_i$) such that $h_1
h_2 \cdots h_k =1$. Then, the trace $Tr$ of the component $U_{h_1,
\dots  ,h_k}$  gives a gauge invariant object which can be
interpreted as the finite group analog of a Wilson loop. When the
volume form $vol=\theta^{i_1} \we \cdots \we \theta^{i_p}$ is
central ($[vol,f]=0, \forall f \in Fun(G)$), then also the
$p$-form $U^p$ is gauge invariant.
 \sk
 \noindent {\bf Note 3.} The gauge invariant
action (\ref{YMaction}) on the finite group lattice is easily
generalized to the case of higher order field strengths:
 \eq
B = B_{h_1, \dots h_k} \theta^{h_1} \otimes \dots \otimes
\theta^{h_k} \; ,
 \en
 transforming as $B \rightarrow T \, B \, T^{-1}$, or
 \eq
 B_{h_1,\dots,h_k} \longrightarrow T \,\,
B_{h_1,\dots,h_k} \,\, [{\cal R}_{h_1, \dots , h_k} T^{-1}] \; ,
 \en
Then
 \eq
  \sum B_{h_1,\dots,h_k} \, B^{~\dagger}_{h_1,\dots,h_k} \longrightarrow T
\, \sum B_{h_1,\dots,h_k} \, B^{~\dagger}_{h_1,\dots,h_k} \, T^{-1}
\; ,
 \en
 If $B$ satisfies $B^\dagger = -B$, then
 $B^{~\dagger}_{h_1...h_k}=(-1)^{k-1}\Rcal_{h_1...h_k}
B_{ad(h_2\ldots h_k)^{-1}h_1^{-1},\,ad(h_3\ldots
h_k)^{-1}h_2^{-1}\!,\, \ldots ,h_k^{-1}}$
 and [use (\ref{dualitypairing})]
 \eq
 B_{h_1,\dots,h_k} \, B^{~\dagger}_{h_1,\dots,h_k}=~ - ~<B,B>
 \en
 The gauge invariant analogue of (\ref{YMaction}) is
 \eq
 A = \sum_G \sum_{h_i \neq e}
 Tr \left(  B_{h_1,\dots,h_k} \,  B_{h_1,\dots,h_k}^{~\dagger}
 \right)= ~
 - ~ \sum_G  Tr ~<B,B>.
 \label{BBaction}
 \en
In particular, setting $B = U^k$, where $U$ is the link 1-form,
and taking into account the condition (\ref{UArealitycomp}), the gauge
invariant action (\ref{BBaction}) is a sum over special Wilson
loops as described in Note 2.


\sect{Duality Rotations}

In this Section we consider an ``abelian'' $U(1)$ gauge theory on
an abelian finite group space. We choose a four dimensional
differential calculus on this finite group. The $U(1)$ gauge
theory is ``abelian'' in the sense that the field strength
satisfies the Bianchi identities $dF=0$, in particular this holds
 if $F=dA$. In this case {\sl infinitesimal} gauge
transformations read $\delta A=d\lambda$, where $\lambda$ is a
gauge parameter, see for ex. the second ref. in \cite{majidetal}.
As in standard electromagnetism the field strength $F$ is
therefore invariant under gauge transformations.

It is remarkable that also for this discrete (noncommutative)
version of electromagnetism one can consider electric-magnetic
duality rotations (for duality rotations in noncommutative
geometry where the noncommutativity is given by a $*$-product we
refer to \cite{AschieriBI} and references therein). Following
\cite{GZGBZ}, see also the nice review \cite{KT}, in this section
we obtain a consistency condition for an (in general nonlinear)
electromagnetic theory to admit duality rotations; we also show
that the equations of motions (EOM) of Maxwell theory admit
electric-magnetic rotations. As far as we know this is the first
example of duality rotations on a lattice.
 \sk
  Let us consider a
$4$-dimensional ${}^*$-calculus on $G$ with one-forms $\th^u,
\th^{u^{-1}}, \th^v,\th^{v^{-1}}$ where $u,u^{-1},v,v^{-1}$ are
the four different elements of $G$ that determine the calculus. It
follows that the metric
$g^{h,k}=<\th^h\,,\,\th^{k}>=-\de^h_{k^{-1}}$ has determinant $\det
g=1$. This is an Euclidean metric as one can check expressing it
in a basis of real one-forms. We also notice that since the finite
group is abelian, the epsilon tensor $\epsi^{hh'gg'}$ is the usual
completely antisymmetric tensor (with $\epsi^{uu^{-1}vv^{-1}}=1$).

Consider an action $S$ that depends on the gauge field $A$ only
through the field strength $F$. The EOM are obtained by varying
$S$ with respect to $A_g$. We have \eqa \de_AS&\equiv&\sum_{p\in
G}{\de S\over \de A_g(p)}\de A_g(p)= {1\over 2}\sum_{p,q\in G}{\de
S\over \de F_{h,h'}(q)}{\de F_{h,h'}(q)\over \de
  A_g(p)}\de A_g(p)
=\sum_{q\in G}{\de S\over \de F_{h,h'}(q)}t_h\de
A_{h'}(q)\nonumber\\ &=&\sum_{q\in G}-t_h({\de S\over \de
F_{h,h'}(q)}) R_{h}\de A_{h'}(q)= \sum_{q\in G}-t_h(R^{-1}_{h}{\de
S\over \de F_{h,h'}(q)}) \de A_{h'}(q)\nonumber \ena where in the
second equality we used that $F_{h,h'}=t_h A_{h'}-t_{h'}A_h$ is
antisymmetric, while in the third  we have integrated by parts via
the Leibniz rule (\ref{tgLeibniz}) $t_g(f h) =t_gf\, R_g h +f t_g
h$ and then used Stokes' theorem: $\sum_{q\in G}t_g(f)(q)=
\sum_{q\in G} (R_g f(q) -f(q))=0$. The EOM are therefore
 \eq
0={\de S\over \de A_{h'}}=-t_h R^{-1}_h {\de S\over \de F_{h,h'}}
~~~~~~~\mbox{(no sum on $h'$)} \label{EOM}
\en
\noi or, applying $R^{-1}_{h'}$ (so that the argument of $t_h$ is
antisymmetric in the indices $h,h'$):
 \eq
  t_h (R^{-1}_{hh'}{\de
S\over \de F_{h,h'}} )=0~.~~~~~~~\mbox{(no sum on $h'$)}
\en
We then define \eq \tilde{G}^{h,h'}\equiv {1\over 2}
\epsi^{hh'gg'}G_{g,g'}\equiv R^{-1}_{hh'}{\de S[F]\over \de
F_{h,h'}} \label{Gtilde}
\en
where we have written $S[F]$ in order to stress that this
definition does not depend on the relation $F=dA$. Using
(\ref{Gtilde}) the EOM and the Bianchi identities $dF=0$ have the
same structure
 \eqa
  t_h \tilde{G}^{h,h'}=0~~,\label{EQ1}\\ t_h
\tilde{F}^{h,h'}=0~~,\label{EQ2}
 \ena
 \noi  where
$\tilde{F}^{h,h'}\equiv {1\over 2} \epsi^{hh'gg'}F_{g,g'}$.
 In summary, starting with the action $S$ we obtained the EOM
 (\ref{EQ1}) with $F=dA$. We now relax the condition $F=dA$
and consider the more general theory given by the EOM
 (\ref{EQ1}) and  (\ref{EQ2}) with $\tilde{G}$ given
in (\ref{Gtilde}) and where {\sl now} S[F] is seen just as a
function of $F$, with $F$ an arbitrary antisymmetric tensor (not
the differential of $A$). It is this theory that possibly admits
duality rotations. More precisely we show that if $S[F]$ satisfies
(\ref{thecondition}) then we have $SO(1,1)$ duality rotations. As
in \cite{GZGBZ} (\cite{KT}) we observe that under the
infinitesimal $GL(2)$ rotation $G\rightarrow G+ \Delta
G,~F\rightarrow F+ \Delta F$ given by \eq
 \Delta \left( \matc G \\
F \emat \right)
=
\left( \matcc \mbox{A} & \mbox{B} \\ \mbox{C} & \mbox{D} \emat
\right) \left( \matc G \\ F \emat \right)~\label{FGtrans}
\en
equations (\ref{EQ1}), (\ref{EQ2}) are mapped into themselves.

\noi For ease of notations, in what follows we'll write $F_{hk}$
instead of $F_{h,k}$, and similarly for ${\tilde F}, G, {\tilde
G}$.

Consistency of (\ref{FGtrans}) with (\ref{Gtilde}), i.e.
 \eq
\tilde{G}^{gg'}+\Delta\tilde{G}^{gg'}=R^{-1}_{gg'} {\de S[F+\Delta
F]\over \de (F_{gg'}+ \Delta F_{gg'})} \label{constr}
\en
constrains the action $S[F]$ and the rotation parameters in
(\ref{FGtrans}). In order to simplify the expression of the
constraint (\ref{constr}) we write
 \eqa
  {\de S[F+\Delta F]\over
\de (F_{gg'}+ \Delta F_{gg'})} &=&\sum_{q\in G} {1\over 2}{\de
S[F+\Delta F]\over \de F_{hh'}(q)} {\de F_{hh'}(q)\over \de
(F_{gg'}+ \Delta F_{gg'})}\nonumber\\ &=&{\de S[F+\Delta F]\over
\de F_{gg'}}- \sum_{q\in G} {1\over 2}{\de S[F]\over \de
F_{hh'}(q)} {\de (\Delta F_{hh'})(q)\over \de F_{gg'}}\nonumber
\ena
 \noi then, using (\ref{FGtrans}) and (\ref{Gtilde}),
condition (\ref{constr}) reads
 \eqa A R_{gg'}\tilde{G}^{gg'}+&&\!\!\!\!\!\!\!\!\!\!\!\! B
R_{gg'}\tilde{F}^{gg'}\nonumber \\ &=&{\de (S[F+\Delta
F]-S[F])\over \de F_{gg'}}-{1\over 2}C \sum_{q\in G} {\de
S[F]\over \de{F}_{hh'}(q)} {\de G_{hh'}(q)\over{\de
F_{gg'}}}-D{\de S[F]\over \de
  F_{gg'}}\label{ABCD}\\
&=& {\de (S[F+\Delta F]-S[F])\over \de F_{gg'}} -{1\over 2}C
\sum_{q\in G} \tilde{G}^{hh'\!}(q) R^{-1}_{hh'}{\de
G_{hh'}(q)\over{\de F_{gg'}}} -D{\de S[F]\over \de
F_{gg'}}\nonumber
 \ena
  where in the last line we used the
invariance of the integral $\sum_{q\in G}$ under the translation
$R^{-1}_{hh'}$. In order to further simplify this expression we
observe that \eq {\de\over \de F_{gg'}} \sum_{q\in
G}F_{hh'\!}(q)R_{hh'}\tilde{F}^{hh'}(q)= 4 R_{gg'}
\tilde{F}^{gg'}\label{FFTilde}
\en
Proof: Use invariance of the integral $\sum_{q\in G}$ under the
translation $R^{-1}_{hh'}$ and then notice that
$R^{-1}_{hh'}\epsi^{hh'gg'}=R_{gg'}\epsi^{hh'gg'}$ since
$h,h,g,g'$ run over the $4$ group elements $u,u^{-1},v,v^{-1}$.

Similarly one has $\de(\sum_{q\in G}
G_{hh'}(q)R_{hh'}\tilde{G}^{hh'}(q)) ={2} \sum_{q\in
G}R_{hh'}\tilde{G}^{hh'}\!(q){}^{\,}\de G_{hh'}\!(q)$ and
therefore \eq {\de\over \de F_{gg'} } \sum_{q\in G}
G_{hh'}(q)R_{hh'}\tilde{G}^{hh'}(q)= 2\sum_{q\in G}
\tilde{G}^{hh'}(q)R^{-1}_{hh'}{\de
  G_{hh'}(q)\over{\de F_{gg'}}}\label{GGtilde}
\en
Next we substitute (\ref{FFTilde}) and (\ref{GGtilde}) in
(\ref{ABCD}) and factorize out the functional derivative $\de\over
\de F_{gg'}$. We thus arrive at the equivalent condition (the
constant term $(A+D)S[F\!=0]$ is obtained by observing that when
$F=0$ also $G=0$) \eq S[F+\Delta F]-S[F] -{B\over 4}  \sum_G
F_{hh'}R_{hh'}\tilde{F}^{hh'} -{C\over 4}  \sum_G
G_{hh'}R_{hh'}\tilde{G}^{hh'}=(A+D)(S[F]-S[F\!=0])
\en
This expression can be further simplified. Use $S[F+\Delta
F]-S[F]={1\over 2} \sum_G {\de S[F]\over \de
  F_{hh'}}\Delta F_{hh'}={1\over 2} C \sum_G
G_{hh'}R_{hh'}\tilde{G}^{hh'} +{1\over 2} D \sum_G
F_{hh'}R_{hh'}\tilde{G}^{hh'}
$
in order to write \eq {C\over 4} \sum_G
G_{hh'}R_{hh'}\tilde{G}^{hh'}-{B\over 4} \sum_G
F_{hh'}R_{hh'}\tilde{F}^{hh'}= (A+D)(S[F]-S[F\!=0])-{D\over 2}
\sum_G F_{hh'}R_{hh'}\tilde{G}^{hh'}\label{eqmpar}
\en
Then, assuming that the Lagrangian is parity even, the parity even
and parity odd terms in (\ref{eqmpar}) must separately vanish and
we obtain
 \eqa C\sum_G G_{hh'}R_{hh'}\tilde{G}^{hh'}=B
\sum_G F_{hh'}R_{hh'}\tilde{F}^{hh'}\label{Dualityrot1}\\
(A+D)(S[F]-S[F\!=0])={D\over 2}  \sum_G F_{hh'}{\de S[F]\over \de
F_{hh'}}\label{Dualityrot2}
 \ena
  \sk
   We now show that linear
electromagnetism satisfies (\ref{Dualityrot1}) and
(\ref{Dualityrot2}) with $B=C$ and $A=D$, i.e. that linear
electromagnetism admits the duality rotation group $SO(1,1)$ times
the group of scale transformations (corresponding to $A=D\not=0$).

Let $$ S_{EM}= -{1\over 4}\sum_G F_{hh'}F^\dagger_{hh'}= {1\over
4}\sum_G F_{hh'}R_{hh'}F_{kk'}g^{hk}g^{h'k'}~, $$ then
$\tilde{G}^{hh'}=F_{ll'}g^{lh} g^{l'h'}\equiv F^{hh'}$, and
therefore $G_{hh'}=\tilde{F}^{ll'}g_{lh}g_{l'h'}\equiv
\tilde{F}_{hh'}$.\footnote{ We recall that det$g=1$ so that
 the Hodge operator $\,\tilde{}$ squares to the
  identity and not to minus the identity as for Minkowski space.
This is why in Euclidean space the duality group is $SO(1,1)$
while in Minkowski space it is $SO(2)$.} Since
$R^{-1}_{hh'}\epsi^{hh'gg'}=R_{gg'}\epsi^{hh'gg'}$, we have
$F_{hh'}R_{hh'}\tilde{F}^{hh'}=\tilde{F}_{hh'}R_{hh'}{F}^{hh'}$
and we conclude that (\ref{Dualityrot1}) is satisfied iff $C=B$.
Similarly ${1\over 2} D \sum_G F_{hh'}{\de S[F]\over \de F_{hh'}}
={2} D S_{EM}[F]$  so that (\ref{Dualityrot1}) is satisfied iff
$A=D$.

For a nonlinear theory $S[F]$, that in the weak field limit
reproduces $S_{EM}[F]$ (i.e. $S[F]=S[F\!=0]+S_{EM}[F]+O(F^4)$),
condition (\ref{Dualityrot2}) is equivalent to
 \eq
D (S[F]-S[F\!=0]-{1\over 4}  \sum_G F_{hh'}{\de S[F]\over \de
F_{hh'}})=0.
 \en
 But $S[F]$ is not homogeneous in $F$ and therefore
(\ref{Dualityrot2}) is satisfied iff $A=D=0$. We conclude that a
nonlinear electromagnetic theory, that has definite parity and
that in the weak field limit reproduces linear electromagnetism,
admits an $SO(1,1)$ duality rotation group if and only if \eq
\sum_G G_{hh'}R_{hh'}\tilde{G}^{hh'}= \sum_G
F_{hh'}R_{hh'}\tilde{F}^{hh'}\label{thecondition}~.
\en

\vspace{0.5cm}
 \noindent
  {\bf Note 1.} In Section 6  we present two Born-Infeld type
actions on finite groups with four independent tangent vectors
(four dimensions). The action (\ref{BIaction}) does not satisfy
condition (\ref{thecondition}). Whether the action
(\ref{BIazione}) in four dimensions admits $SO(1,1)$ duality
rotations remains to be determined.


\sect{Kaluza-Klein gauge theory on $M^D \times G$}

We extend here the results of Section 3 to include spaces of the
type $M^D \times \mbox{finite} ~G$. This extension is
straightforward (see also \cite{tmr}), and allows us to apply
Kaluza-Klein techniques when the internal space is a discrete
finite $G$ ``manifold".

We'll use the letter $x$ for the $M^D$ coordinates, and $y$ for
the $G$ coordinate functions.  A basis of 1-forms on $M^D \times
G$ is given by $dx^A=\{dx^\mu, \theta^g\}$, with ${x^\mu}^*=x^\mu$
and
 \eqa
 & &  dx^\mu \we \theta^g = - \theta^g \we dx^\mu \\
 & & (dx^\mu \we \theta^g)^* = - (\theta^g)^* \we
 (dx^\mu)^* =  \theta^{g^{-1}} \we dx^\mu =
 - dx^\mu \we \theta^{g^{-1}}\\
 & & dx^{A_1} \we ...\we dx^{A_{D+p}} \equiv \epsilon^{A_1...A_{D+p}}
 ~vol(M^D \times G) \\
 & & dx^{\mu_1}\we ... \we dx^{\mu_D} \we
  \theta^{g_1} \we ...\we \theta^{g_p} = \epsilon^{\mu_1 \cdots
 \mu_D} \epsilon^{g_1...g_p}~vol(M^D) \we vol(G)\\
 & & \int_{M^D \times G} f(x,y) \equiv \int_{M^D} d^Dx \left(
 \int_G f(x,y)~ vol(G) \right)\\
 & & <dx^{\mu_1} \we ...\we dx^{\mu_k}, dx^{\nu_1} \we ...\we
 dx^{\nu_k}> = (k!)^2~\de^{\mu_1 ...\mu_k}_{\nu_1 ...\nu_k}, \\
 & & <dx^\mu  \we \theta^g,dx^\nu \we \theta^h>=-2 \de^\mu_\nu
 \de^g_{h^{-1}}
 \ena
\noi where $vol(M^D \times G)= vol(M^D) \we vol(G)=d^Dx ~vol(G)$,
$\epsilon^{\mu_1 ... \mu_D}$ is the usual Levi-Civita tensor and
$\epsilon^{g_1...g_p}$ has been defined in (\ref{vol}). The
normalization of exterior products is such that, for example,
$dx^\mu \we dx^\nu = dx^\mu \otimes dx^\nu - dx^\nu \otimes
dx^\mu$ and similar for $dx^\mu \we \theta^g$.

 The gauge potential 1-form  $A(x,y)$ is then expanded as:
 \eq
 A(x,y) = A_{\mu}(x,y)~dx^\mu + A_g (x,y)~\theta^g
 \en
 The gauge variation $A' = -(dT)T^{-1}+TAT^{-1}$ becomes the usual
 one for the components $A_\mu$, and the one given in
 (\ref{gaugeAcomp}) for $A_g$, with $T$ depending
  on $x$ and $y$. Thus $A_\mu$ belongs to the Lie
 algebra of $\Gcal$, whereas $A_g$ belongs to the group algebra of
 $\Gcal$.

Defining as usual the field strength as $F=dA+A\we A$ and its
components as
 \eq
 F \equiv {1\over 2} F_{\mu\nu} ~ dx^\mu \we dx^\nu +  F_{\mu k}~ dx^\mu \we
 \theta^k +  F_{h,k} ~\theta^h \otimes \theta^k
 \en
 \noi we find
 \eqa
 & & F_{\mu\nu}=\partial_\mu A_\nu - \partial_\nu A_\mu +
  A_\mu A_\nu - A_\nu A_\mu \\
 & &  F_{\mu k} = \partial_\mu U_k + A_\mu U_k - U_k (\Rcal_k
 A_\mu) \equiv D_\mu U_k  \label{Fmug} \\
 & & F_{h,k}=U_h (\Rcal_h U_k)-U_k (\Rcal_k U_{k^{-1}hk})
 \ena
 where the link field is defined as (cf. Section 3) $U_g(x,y)=
 1+A_g (x,y)$. The gauge transformation $F'(x,y)=T(x,y)F(x,y)T^{-1}(x,y)$ implies:
 \eqa
 & & F_{\mu\nu}'(x,y)=T(x,y) F_{\mu\nu}(x,y) T^{-1}(x,y)\\
  & & F_{\mu k}'(x,y)= T(x,y) F_{\mu k}(x,y) \Rcal_k T^{-1}(x,y)
  \\
  & & F_{h,k}'(x,y)=T(x,y)  F_{h,k}(x,y) \Rcal_{hk} T^{-1}(x,y)
  \ena
  {} From the antihermiticity $A^\dagger = -A$ (or $U^\dagger = -U$),
  one finds, as in Section 3, $F^\dagger = -F$. In components:
  \eq
  F^\dagger_{\mu\nu}=-F_{\mu\nu},~~F^\dagger_{\mu k}= \Rcal_k
  F_{\mu k^{-1}},~~F^\dagger_{h,k}=-\Rcal_{hk} F_{k^{-1}h^{-1}k,k^{-1}}
  \en
  Then the Yang-Mills action on $M^D \times G$ can be expanded as
   follows:
  \eqa
  & &A_{YM} = -\int_{M^D \times G} Tr~F \we *F =  -\int_{M^D \times G}
  Tr ~<F,F> vol(M^D \times G) = \nonumber\\
  & & = -\int_{M^D} d^Dx \sum_{G}~
  Tr~[F_{\mu\nu} F_{\mu\nu}-2 F_{\mu k} F_{\mu
  k}^\dagger-F_{h,k} F_{h,k}^\dagger]= \nonumber \\
  & & = \int_{M^D} d^Dx \sum_{G}~ Tr~[-F_{\mu\nu}
  F_{\mu\nu}+ 2 D_\mu U_k (D_\mu U_k)^\dagger + \nonumber\\
  & & ~~~~~~~~~~~~~~~~~~~~~~~~
    + 2 U_k U^\dagger_k U_h U^\dagger_h-2 U^\dagger_k U_h
 (\Rcal_h U_k)(\Rcal_k U^\dagger_{k^{-1}hk})] \label{AYMU}
   \ena
   Note that this action is real, and describes
  a Yang-Mills theory in $D$ dimensions minimally coupled to
  the scalar fields $U_g$.


\sect{Kaluza-Klein gauge theory on $M^D \times Z_N$}


In this section we first study the geometry of $Z_N$ equipped with
a bicovariant calculus and a ${}^*$-conjugation, and consider the
limit $N\rightarrow \infty$ i.e. $Z_N\rightarrow S_1$. It is then
easy to generalize the results to $M^D\times Z_N$, consider the
Yang-Mills action on this space and understand its $N\rightarrow
\infty$ limit.

\subsection{${}^*$-bicovariant calculus on $Z_N$}

\noindent Let $u$: $u^j u^k=u^{j+k},~ u^N=u^0=e$ be the generator
of the cyclic group $Z_N$. A  basis of  functions on $Z_N$ is
given by $x^{u^j} = \{x^e,x^u,x^{u^2},...,x^{u^{N-1}} \}$. It is
convenient to use a basis of functions that reproduce the algebra
of the $Z_n$ elements $u^j$. This basis is given by
\cite{DMGcalculus} $y^j \equiv \sum^{N-1}_{k=0} q^{jk} x^{u^k}$,
where $q \equiv e^{{2\pi i\over N}}$. Thus $y^j y^k = y^{j+k}$,
$y^0=I$. For example $y^1=y$ is given by
 \eq
  y=x^e+qx^u+q^2 x^{u^2}+...q^{n-1} x^{u^{n-1}} \label{newy}
 \en
 having the values:
 \eq
 y(u^k)=q^k= e^{2\pi i{k\over N}}
 \en
 on the $N$ points of $Z_N$.
Using $\sum_{j=0}^{N-1} q^{kj}=N~ \delta_{k,0}$ one finds the
inverse transformation: $x^{u^j}={1\over N} \sum_{k=0}^{N-1}
q^{-jk} y^k$.

The cyclic group $Z_N$ can be seen as a discrete approximation of
the circle $S_1$ of radius $R$. Let $0 \leq x \leq 2 \pi R$ be the
$S_1$ coordinate on the circle (not to be confused with the basis
functions $x^{u^j}$ of $Z_N$) : then the points of $Z_N$ have
coordinates $x_k =x(u^k)= 2 \pi R k/N$ on the circle, and their
corresponding $y$ values are $y(x_k) = e^{i {x_k\over R}}$. In the
limit $N \rightarrow \infty$ these points fill the whole circle,
and the $x_k$ become the continuous values of the $S_1$ coordinate
$x$.

  Among the many differential calculi on $Z_N$ the most discussed in the
literature \cite{DMGcalculus,gravfg,CasPag} is the one where only
$\theta^u=-\theta^e\not= 0$. This calculus  is not (for $N\not=2$)
a ${}^*$-calculus: $(df)^*\not= d(f^*)$. The compatibility of a
differential calculus with the ${}^*$-product is essential if we
want to consider the calculus on $Z_N$ as a discretization of the
{\sl real} differential calculus on $S_1$. Indeed on a circle of
radius $R$ we have $(df)^* = d(f^*)$, where ${}^*$ denotes complex
conjugation, and $f=\sum_nf_n y^n(x)=\sum_nf_n e^{i n x/R}$,
$f_n\in\,${\bf C}, $x^*=x$, i.e. $y^*(x)=y^{-1}(x)$. In the
following we therefore choose the ${}^*$-bicovariant calculus
generated by the $1$-forms $\th^u$ and $\th^{u^{-1}}$. This is the
minimal ${}^*$-bicovariant calculus with  $\th^u$ nonzero. In
fact, from $y^*=y^{-1}$ and  $(hdf)^*=d(f^*)\, h^*$ we obtain
${\th^u}^{\,*}=-\th^{u^{-1}}$.\footnote{If $N$ is odd there is no
$1$-dimensional $*$-bicovariant calculus and the $2$-dimensional
$*$-bicovariant calculus defined by  $\th^u$ and $\th^{u^-1}$ is
the most natural. If $N$ is even on the contrary there exist a
unique $1$-dimensional $*$-bicovariant calculus, it is generated
by the $1$-form $\th^{u^{N/2}}$ that is pure imaginary
${\th^{u^{N/2}}}^{\,*}=-\th^{u^{-N/2}}=-\th^{u^{N/2}}$; for
$N\rightarrow \infty$ however this calculus does not lead to the
standard calculus on $S_1$.}

In order to gain a better insight about the geometry of $Z_N$ it
is convenient to consider the real and closed forms $$ \al=-i R
y^{-1}dy~,~~\beta= i R y dy^{-1}~~; $$ in the $N\rightarrow
\infty$ limit, requiring that the exterior differential $d$
becomes the commutative one, we find $\al=\beta=dx$, $0\leq x\leq
2\pi R$. For finite $N$, $\al$ and $\beta$ are linearly
independent, and recalling (\ref{dxastheta}) and (\ref{newy}) we
obtain
 \eq
  \al=iR(\zeta-\zeta^*)~,~~\beta=iR(q^{-1}\zeta
  -q\zeta^*) \lb{albe}
 \en
where we have defined
 \eq
 \zeta=(1-q)\th^{u}~,~~
 \zeta^*=(q^{-1}-1)\th^{u^{-1}}~.
 \en
{}From (\ref{albe}) we see that $\zeta$, $\zeta^*$ are $\not=0$ in
the $N\rightarrow \infty$ limit and therefore $\th^{u}$ and
$\th^{u^{-1}}$ are ill defined in this limit.

\sk
 \noi {\bf Metric}

\noindent Using $y \theta^{u}=q^{-1} \theta^{u} y~,\,~y
\theta^{u^{-1}}=q\theta^{u^{-1}}y\,$ it is easy to see that, up to
normalization, there exists a unique metric (bimodule pairing) on
the space of $1$-forms such that on left-invariant $1$-forms it is
symmetric, and satisfies the properties in
(\ref{metricproperties}). The normalization, compatible with the
$N\rightarrow \infty$ limit, is naturally fixed requiring
$<\!\al\,,\,\al>=1$.  Explicitly we have \eqa
g^{\zeta\zeta^*}=<\zeta,\zeta^*>={1\over{2 R^2}}~,~~
g^{\zeta^*\zeta}=<\zeta^*,\zeta>={1\over{2 R^2}}\\
g^{\zeta\zeta}=<\zeta,\zeta>=0~,~~
g^{\zeta^*\zeta^*}=<\zeta^*,\zeta^*>=0\label{metrica}
 \ena
 \noi  We then also have $<\beta,\beta>=1$. As in (\ref{starmetric}), this
pairing is compatible with the ${}^*$-conjugation $ <\rho,\sigma
>^*=<\sigma^*,\rho^*>\,.
$
Having the metric $g$, we can find the $1$-form $\gamma$ that is
orthogonal to $\al$ and has unit length,
$<\al\,,\,\gamma>=0\,,~<\gamma\,,\,\gamma>=1$. This form is given
by
 \eq
  \gamma=-R(\zeta+\zeta^*)=i{{q+q^{-1}}\over{q-q^{-1}}}
\al-{{2i}\over{q-q^{-1}}} \beta \lb{gamma}
\en
Notice that $\gamma^*=\gamma$. In the $N \rightarrow \infty$ limit
$\al,\be,\ga$ are well defined ($\al$ and $\be$ become $dx$, and
$\ga$ can be checked from (\ref{gamma}) or from $<\ga,\ga>=1$ to
remain finite). Then from (\ref{gamma}) and (\ref{albe}) we see
that both $\zeta$ and $\zeta^*$ are well defined (besides being
$\not=0$) in the $N\rightarrow \infty$ limit, while $\th^u$ and
$\th^{u^{-1}}$ diverge as $1/(q^{-1}-1)$. One also has
$t_u/(q^{-1}-1)\rightarrow iR{\partial\over {\partial x}}$ for
$N\rightarrow \infty$.

The pairing $<\!\!~,\!\!~>$ can be generalized to act on the space
of $2$-forms. The space of left-invariant $2$-forms is one
dimensional because the wedge product for abelian groups is the
commutative one (e.g. $\zeta\wedge\zeta^*=\zeta\otimes
\zeta^*-\zeta^*\otimes\zeta$). Any $2$-form can be written as $f~
vol$ where $vol$ is the volume form associated to the metric:
$$vol=\al\wedge \gamma=-2iR^2\zeta\wedge \zeta^*~.$$ Up to a
normalization the ($*$-bimodule) pairing $<~,~>$ is uniquely
defined. As in the commutative (continuous) case we choose here
the normalization such that  $$<vol,vol>=1~.$$ Notice that $vol$
is central, therefore the space of $2$-forms has no
noncommutativity. Notice also that
$vol^*=(-1)^{|\al||\gamma|}\gamma^*\wedge\al^*=vol$.

\sk

\noindent {\bf $*$-Hodge operator}

\noindent The $*$--Hodge operator is defined as in
(\ref{defhodge}). We have
 \eq
 *\zeta=i\zeta~~,~~
  *\zeta^*=-i\zeta^*~.
   \en

\sk \noindent {\bf Integral}

\noindent The Haar measure $h$ on $Z_N$ is given by
 \eq
h(f)=\sum_{k=0}^{N-1} f(u^k)
\en
Integration on $2$-forms is then simply given by \eq \int f
vol\,={2\pi R \over N} h(f)\lb{integral}
\en
the normalization is chosen such that in the $N\rightarrow \infty$
limit we obtain the usual integral over the circle $S_1$ of radius
$R$. It is not difficult to check that the integral
(\ref{integral}) is cyclic, that it satisfies Stokes' theorem and
that it is real. For any $p$- and $(2-p)$-form $\omega^{(p)},
\omega^{(2-p)}$, we have \eq \int
\omega^{(p)}\wedge\omega^{(2-p)}=(-1)^{p(2-p)}\int
\omega^{(2-p)}\wedge\omega^{(p)} ~~~~~~~\mbox{(cyclicity)}
\lb{cyclic}
\en
\eq \int d\omega^{(1)}=0
\en
\eq \left(\int\omega^{(2)}\right)^*=\int{\omega^{(2)}}^*
\label{reale}
\en
\subsection{${}^*$-bicovariant calculus on $M^D \times Z_N$}
It is straightforward to generalize the results of the previous
subsection to the $M^D\times Z_N$ case. A basis of $1$-forms is
given by $\{dx^A\}=\{dx^\mu,\al,\gamma \}$, ${x^\mu}^*=x^\mu$. The
metric $g^{AB}$ is block diagonal $g^{\mu\nu}\not=0$,
$g^{\mu\al}=g^{\mu\gamma}=g^{\al\gamma}=0$,
$g^{\al\al}=g^{\gamma\gamma}=1\,.$ The volume element is
$vol=\sqrt{|{\rm{det}} g_{AB}|}\,~ dx^0\wedge dx^1\wedge\ldots
dx^{D-1}\wedge\al\wedge\gamma$, (where
$g^{AC}g_{CB}=\delta^A_{\,B}$). In the  $\{dx^A\}$ basis the
epsilon tensor, defined by
 $$dx^{A_1}\wedge dx^{A_2}\wedge\ldots
dx^{A_{D+2}}= \varepsilon^{A_1 A_2\ldots A_{D+2}} vol$$ is the
classical one. The integral
 $$ \int_{M^D\times Z_N}\!\!
f(x^\mu,y)\, vol\,\equiv\,\int_{M^D}\! \sqrt{|\det g_{\mu\nu}|}~
d^D x \,\left(\int_{Z_N} f(x^\mu,y)\al\wedge\gamma\right) $$ is
again cyclic, real  and satisfies Stokes' theorem.

The $*$-Hodge operator is still given by (\ref{defhodge}), and
satisfies the left- and right-linearity property
(\ref{linearityhodge}) and the compatibility with the
${}^*$-conjugation (\ref{compatibilehodge}). The normalization of
the pairing can be chosen such that the Hodge operator squares to
$\pm id$.

\sk
\subsection{Yang-Mills action}
The  Yang-Mills action on $M^D\times Z_N$ with gauge group ${\cal
G}=U(M)$ is given by \eq A_{YM}=  -\int_{M^D\times Z_N}Tr(F\wedge
*F) \label{YMazione}
\en
where the trace is over $M\times M$  matrices.
Positivity and gauge invariance of this action are shown in
Section 4. In order to write the Yang-Mills action in terms of $F$
components we first obtain the commutation relations \eqa
 \al y^n = {q^n+q^{-n}\over 2} y^n \al -i {q^n-q^{-n}\over
2}y^n \gamma \equiv {\tilde y}_n\al -i{\check y}_n \gamma \\
\gamma y^n = {q^n+q^{-n}\over 2} y^n \gamma +i {q^n-q^{-n} \over
2}y^n \al ={\tilde y}_n\gamma +i{\check y}_n \al
 \ena
  we then expand $F_{AB}(y)$ as
   \eq F_{AB}(y) =F_{AB\,(k)}y^k{1\over \sqrt{2\pi R}} ~~
\en
and define
 \eq F_{AB}(\tilde{y})\equiv F_{AB\,(k)}\tilde
{y}_k{1\over \sqrt{2\pi R}} ~~,~~~ F_{AB}(\check{y})\equiv
F_{AB\,(k)}\check {y}_k{1\over \sqrt{2\pi R}} ~~.
\en
Finally the action reads
 \eq
\begin{array}{lrl}
\displaystyle A_{YM}&=& \int_{M^D\times Z_N}Tr [
F_{\mu\nu}F^{\mu\nu} + F_{\al\gamma}F^{\al\gamma} + 2
F_{\al\mu}(y)F^{\al\mu}(\tilde{y}) +\\ & &~~~~+ 2F_{\gamma\mu}(y)
F^{\gamma\mu}(\tilde{y}) -2i F_{\al\mu}(y)F^{\gamma\mu}(\check{y})
+2i F_{\gamma\mu}(y)F^{\al\mu}(\check{y})\,] \,vol\\[1.1em] &=&
\sum_k\int_{M^D}\sqrt{|{\rm{det}}g_{\mu\nu}|}\, Tr [
F_{\mu\nu\,(k)}F^{\mu\nu}_{~~(-k)} + F_{\al\gamma\,(k)}
F^{\al\gamma}_{~~(-k)} +2q^k F_{\al\mu\,(k)}F^{\al\mu}_{~~(-k)}\\
& &~~~~+2q^k F_{\gamma\mu\,(k)}F^{\gamma\mu}_{~~(-k)} -2i
(q^k-q^{-k})F_{\al\mu\,(-k)}F^{\gamma\mu}_{~~(k)}\,]
\end{array} \label{YMonMZ}
\en
 \sk
 \noi where in the second equality we have integrated on $Z_N$
using (\ref{integral}) and $\sum_{j=0}^{N-1} q^{kj}=N~
\delta_{k,0}$.

 In the $N\rightarrow \infty$ limit $F_{AB(k)}$ becomes the $k-th$
 Fourier mode, and the action (\ref{YMonMZ}) becomes the
dimensional reduction (in the direction $\gamma$) of the usual
Yang-Mills action on $M^D\times S_1\times S_1$, where the first
$S_1$ is in the $\al$ direction and the second $S_1$ is in the
$\gamma$ direction. It is therefore  Yang-Mills theory on
$M^D\times S_1$ coupled to the adjoint scalar $\phi=A_{\gamma}$.

The interesting feature of this action arises for finite $N$: in
this case we have a nontrivial scalar potential term. Using the
link fields $U$ it is given in (\ref{AYMU}). These variables are
not convenient in the present section because they are ill defined
in the $N\rightarrow\infty$ limit.


\sect{Born - Infeld Theory on finite group spaces}


 We recall the continuum $D$-dimensional
  Born - Infeld action for
non-linear electrodynamics \cite{BI} in flat space
 \eq
 A_{BI} =  \int_{M^D} d^Dx \sqrt{ \det ( \delta_{\mu\nu} +
 F_{\mu\nu})
} \label{BIasdet}
 \en
 When $D=4$ this action takes the form (after explicitly computing
 the determinant):
 \eq
 A_{BI} = \int_{M^4} d^4 x~\sqrt{ 1 + \frac{1}{2} \,  F \cdot F +\,
\left(\frac{1}{4} \, F \cdot \tilde{F} \right)^2 } \; ,
\label{BInodet}
 \en
where $\tilde{F}_{\mu\nu} = \frac{1}{2} \,
 {\epsilon}_{\mu\nu\rho\sigma} F_{\rho\sigma}$, the dot in products
means complete index contraction (for ex. $F \cdot F \equiv
F_{\mu\nu} F_{\mu\nu}$), and we consider the euclidean theory.

 The action (\ref{BIasdet}) can be generalized to the nonabelian case:
then $F_{\mu\nu}$ is $\Gcal$ Lie algebra valued, and the
determinant in (\ref{BIasdet}) is not a number any more but
belongs to the universal enveloping algebra of $\Gcal$. We can
define its ``absolute value" $|\det|$ (still belonging to the
enveloping algebra) as the positive square root in
$\sqrt{\det~{\det}^\dagger}$. Any square root of $|\det|$ is
hermitian, so that an overall trace produces a real action:
 \eq
 A_{BI} =  \int_{M^D} d^Dx ~Tr \sqrt{ |\det ( \delta_{\mu\nu} +
 F_{\mu\nu})|} \label{BIasdet2}
 \en
 This is a gauge invariant action for any choice of
 the square root. The trace can be symmetrized so to fix ordering
ambiguities in products of the Lie valued $F_{\mu\nu}$ elements
 (see for ex. \cite{Tseytlin} and included references).

  We address now the problem of formulating BI theory
  (for a nonabelian gauge group $\Gcal$)  on finite group spaces.

 We first consider the special $D=4$ case (\ref{BInodet}) on finite groups:
  since the indices of $F$ run here on 4 values, we take finite groups with four
independent tangent vectors (for ex. $Z_N \times Z_N$ with the
*-bicovariant calculus of Section 6.1). Later we present a
generalization of the action (\ref{BIasdet}) on any finite group
$G$.

As we have done for the Yang-Mills action, we replace the term $F
\cdot F$ by $<F,F> = -  F_{rs} F_{rs}^\dagger$. Using the bimodule
pairing (\ref{metricproperties}), also the quartic term $(F \cdot
\tilde{F})^2$ can be replaced by the gauge covariant expression
 \eq
<F^2,F^2>\equiv <F\we F,F\we F> \label{F2F2}
 \en
 transforming under gauge variations as
 $<F'^2,F'^2>=T<F^2,F^2>T^{-1}$.
 Then the whole action:
 \eq
 A_{BI}= \sum_G Tr \sqrt{ 1 + \frac{1}{2}  \, <F,F> + \,
\frac{1}{16}   \, < F^2, F^2> }  \label{BIaction}
\en
 is gauge invariant, and reproduces in the continuum case the
 Born-Infeld action for nonabelian gauge fields.
\sk
 Let us explore in more detail the structure of $< F^2, F^2>$,
 and its expression in terms of the link field $U_h$.
 The differential 4-form $F^2$
 \eqa
\lb{F2}
 & & F^2 =  U^4 = U_{h_1} \, [ {\cal R}_{h_1} U_{h_2} ] \,
[ {\cal R}_{h_1 h_2} U_{h_3} ] \, [ {\cal R}_{h_1 h_2 h_3} U_{h_4}
] \, A^{h_1 ,h_2 ,h_3 ,h_4}_{h'_1 ,h'_2 ,h'_3 ,h'_4} \theta^{h'_1}
\otimes \theta^{h'_2} \otimes \theta^{h'_3} \otimes \theta^{h'_4}
= \nonumber \\
 & & ~~~~~~~~~~~~= (F^2)_{h_1,h_2,h_3,h_4} \theta^{h_1} \otimes
\theta^{h_2} \otimes \theta^{h_3} \otimes \theta^{h_4} \; ,
 \ena
\noi transforms under gauge variations as:
 \eq
\lb{F2gauge} F^2  \longrightarrow T \, F^2 \, T^{-1} \;\; , \;\;\;
(F^2)_{h_1,h_2,h_3,h_4} \longrightarrow T \,\,
(F^2)_{h_1,h_2,h_3,h_4} \,\, [{\cal R}_{h_1 h_2 h_3 h_4} T^{-1}]
\; .
 \en
 Consider then the quartic term (sum on the indices $h_i$ understood)
 \eq
 \lb{F2F2bis}
(F^2)_{h_1,\dots,h_4} \, (F^2)_{h_1,\dots,h_4}^\dagger
 \en
This term is gauge covariant, i.e. it transforms as $T \cdots
T^{-1}$, and in fact coincides with $<F^2,F^2>$
 [use (\ref{dualitypairing})].
\sk

  \noi {\bf Born-Infeld action on arbitrary finite groups }
  \sk

 The analogue of $\de_{\mu\nu}+F_{\mu\nu}$ becomes simply
 $E_{g,h} \equiv - \de_{g,h^{-1}} + F_{g,h}$, cf. (\ref{metric}),
 and transforms under gauge variations in the same way as $F_{g,h}$:
  \eq
  E'_{g,h}=-\de_{g,h^{-1}} + T F_{g,h} \Rcal_{gh}T^{-1} =
  T (-\de_{g,h^{-1}} +F_{g,h})  \Rcal_{gh}T^{-1}= T E_{g,h}
  \Rcal_{gh}T^{-1} \label{Dtransformation}
  \en
 We need now a gauge covariant definition of determinant for a matrix
 transforming as in (\ref{Dtransformation}), and that possibly reduces
  to the usual determinant in some limit that recovers the
  continuum case. This limit exists for $N \rightarrow \infty$ in
  the case $G = Z_N \times Z_N \times ... \times Z_N$, and indeed
  the definition we propose in the following has this property.
  \sk
  \noi {\sl Lemma:}
  \eq
    \epsilon^{g_1...g_p} \epsilon^{h_1...h_p}~
     \Rcal_{g_1...g_p} \Rcal_{h_1...h_p} =
    \epsilon^{g_1...g_p} \epsilon^{h_1...h_p}~id~~~~\mbox{(no sums on $g$, $h$)}
  \label{Lemma}
   \en
  \noi {\sl Proof:} for any function $f$:
  \eqa
   & &  \epsilon^{g_1...g_p} \epsilon^{h_1...h_p} ~f
   ~\Ncal = <\theta^{g_1} \we ...\we \theta^{g_p},\theta^{h_1} \we ...\we
   \theta^{h_p}> ~f = \nonumber \\
   & & = <\theta^{g_1} \we ...\we \theta^{g_p},
   (\Rcal_{h_1...h_p}f)~
   \theta^{h_1} \we ...\we \theta^{h_p}> = \nonumber \\
   & &(\Rcal_{g_1...g_p}\Rcal_{h_1...h_p}f)~<\theta^{g_1} \we ...\we
\theta^{g_p},\theta^{h_1} \we ...\we
   \theta^{h_p}>= \nonumber \\
   & &  \epsilon^{g_1...g_p} \epsilon^{h_1...h_p}~
   (\Rcal_{g_1...g_p}\Rcal_{h_1...h_p}f) ~\Ncal
   \ena
   proving the Lemma ($\Ncal$ is defined as $<vol,vol>$,
   cf. (\ref{volvol})).
   \sk

  \noi {\sl Proposition :} the determinant of the matrix $E_{g,h}$
  defined as:
  \eqa
  & & {\det}_G E_{g,h }= \nonumber\\
  & & \epsilon^{g_1, \dots, g_p} ~ E_{g_1,h_1} ~ ({\cal
R}_{g_1 h_1} E_{g_2,h_2} ) ~ ({\cal R}_{g_1 h_1 g_2 h_2}
 E_{g_3,h_3}) \dots ({\cal R}_{g_1 h_1 g_2 h_2 \dots g_{p-1}
h_{p-1}} E_{g_p,h_p})
 ~ \epsilon^{h'_1, \dots, h'_p} \equiv \nonumber\\
 & & \equiv  \epsilon^{g_1, \dots, g_p} ~
 E_{g_1,h_1,g_2,h_2,\dots , g_p,h_p}  ~ \epsilon^{h'_1, \dots,
h'_p} \label{detG}
 \ena
  where $h'_n = (g_{n+1} g_{n+2} \dots g_p)^{-1} h_n
(g_{n+1} g_{n+2} \dots g_p)$, transforms covariantly:
 \eq
  {\det}_G E'_{g,h} = T~ {\det}_G E_{g,h} ~T^{-1}
  \label{detGvariation}
  \en
  \noi {\sl Proof:} the quantity defined in the last line of
  (\ref{detG}) transforms as:
  \eqa
 & & E'_{g_1,h_1,g_2,h_2,\dots , g_p,h_p} = T~
 E_{g_1,h_1,g_2,h_2,\dots , g_p,h_p} ~ ({\cal R}_{g_1 h_1 g_2 h_2
 \dots g_{p} h_{p}} T^{-1}) = \nonumber\\
 & & = T ~ E_{g_1,h_1,g_2,h_2,\dots , g_p,h_p} ~
 ({\cal R}_{g_1 g_2 \dots
 g_{p} h'_1 h'_2 \dots h'_{p}} T^{-1}) \label{Ebigtransf}
 \ena
 where we used the definition of the $h'$ indices. Then
 recalling ${\cal R}_{g_1 g_2 \dots
 g_{p} h'_1 h'_2 \dots h'_{p}}=\Rcal_{g_1...g_p}
 \Rcal_{h'_1...h'_p}$ the gauge variation of
 ${\det}_G E_{g,h}$ reads
 \eq
 {\det}_G E'_{g,h}=T~ \epsilon^{g_1, \dots, g_p} ~
 E_{g_1,h_1,g_2,h_2,\dots , g_p,h_p} (\Rcal_{g_1...g_p}
 \Rcal_{h'_1...h'_p} T^{-1})\epsilon^{h'_1, \dots, h'_p}
 =  T~ {\det}_G E_{g,h} ~T^{-1}
 \en
 after using the Lemma (\ref{Lemma}) in the last equality.
 \sk
  For abelian $G$ and constant matrices this determinant coincides
 with the usual determinant, multiplied by $p!$ since we have defined it
 by means of two $\epsilon$ tensors. We could equally well define
 ${\det}_G$ by fixing the order of $g_1...g_p$ to be the one that
 defines the volume form, i.e. $vol = \theta^{g_1} \we ... \we
 \theta^{g_p}$ so that $\epsilon^{g_1...g_p}=1$ disappears
 from the formula and the indices $g_1...g_p$ are not summed any more.

 \sk
 A covariant {\sl hermitian} ``absolute value" of the determinant
  can be defined by choosing the positive square root in:
 \eq
 |{\det}_G (E_{g,h})| \equiv \sqrt{{\det}_G E_{g,h} ({\det}_G
 E_{g,h})^\dagger }\label{RdetG}
 \en
 and can be used to construct a real $\Gcal$-gauge invariant
 Born-Infeld action on the finite group $G$:
  \eq
  A^G_{BI} = \int_G Tr~\sqrt{|{\det}_G ~(-\de_{g,h^{-1}} + F_{g,h})|} ~vol(G) =
   \sum_G Tr~ \sqrt{| {\det}_G~ (- \de_{g,h^{-1}} + F_{g,h})|}
   \label{BIazione}
   \en
   for any choice of square root; a positive definite action is obtained
    by choosing the positive square root.
    Note that for finite groups the
   order of the $F$ factors in the definition of ${\det}_G$ is
   fixed by (\ref{detG}).

 As an example consider $G=Z_N$ with the 2-dimensional differential calculus
 involving the two left-invariant 1-forms $\theta^u$ and
 $\theta^{u^{-1}}$. The matrix $E_{g,h}$ becomes:
 \eq
 \left(
\begin{array}{cc}
0 & - 1 + F_{+-} \\ -1 - F_{+-} & 0
\end{array}
\right) \; ,
 \en
 where $F_{+-} = F_{u,u^{-1}}$. The determinant (\ref{detG})
 in this case coincides with the usual determinant
  \eq
   {\det}_G E_{g,h} =  \det E_{g,h} = - 1 + F_{+-}^2 \;\;\; .
  \en
  A real gauge invariant BI action is then obtained as:
  \eq
  A_{BI} = \sum_{Z_N} Tr ~ \sqrt{ |-1
  + F^2_{+-}|}
 \en
  the positive square root yielding a positive definite action.

\sect{Born - Infeld Theory on  $M^D \times G $}


 It is convenient to extract the $E=g+F$ matrix from the following
 expansion of $F$:
 $$
 F \equiv {1\over 2}F_{\mu\nu} ~dx^\mu \wedge dx^\nu + F_{\mu g}~ dx^\mu \wedge
 \theta^g + F_{g,h} ~\theta^g \otimes \theta^h =
 $$
 $$ = dx^\mu \otimes  F_{\mu\nu} ~ dx^\nu + dx^\mu \otimes
  F_{\mu h}~ \theta^h - \theta^g \otimes ~  {\cal R}_{g^{-1}}
F_{\nu g}~  dx^\nu + \theta^g \otimes {\cal R}_{g^{-1}} F_{g,h} ~
\theta^h = $$
 \eq  = (dx^\mu ~ , ~ \theta^g ) \otimes
\left(
\begin{array}{cc}
 F_{\mu\nu} &  F_{\mu h} \\
- ~ {\cal R}_{g^{-1}} F_{\nu g} & {\cal R}_{g^{-1}} F_{g,h}
\end{array}
\right) \left(
\begin{array}{c}
dx^\nu \\ \theta^h
\end{array}
\right)
 \en
 where we used
 $$
 dx^\mu \wedge \theta^g =
dx^\mu \otimes \theta^g - \theta^g \otimes dx^\mu
 $$
 Then the $E_{A,B}$ ($A=\mu,g,~B=\nu,h$) matrix is given by:
 \eq E_{A,B}=
 \left(
\begin{array}{cc}
g_{\mu\nu} + F_{\mu\nu} \; , &   F_{\mu h} \\ - ~ {\cal
R}_{g^{-1}} F_{\nu g} \; , &  -\delta_{g,h^{-1}} + {\cal
R}_{g^{-1}} F_{g,h}
\end{array}
\right)
\en
 Next we have to define the determinant of a matrix with mixed indices $\mu,g$.
 Again we find a definition that ensures gauge invariance,
  and a correct continuum limit when it exists. Consider the algebraic identity
   \eq
  \left( \begin{array}{cc}
   A & B \\
   C & D \
 \end{array} \right) =
 \left( \begin{array}{cc}
   1 & B \, D^{-1} \\
   0 & 1 \
 \end{array} \right) ~
 \left( \begin{array}{cc}
   A - B \, D^{-1} C & 0 \\
   0 & D \
 \end{array} \right) ~
 \left( \begin{array}{cc}
   1 & 0 \\
   D^{-1} \, C & 1 \
 \end{array} \right)
 \en
  With the usual definition of determinant this implies:
   \eq
 \det  \left( \begin{array}{cc}
   A & B \\
   C & D \
 \end{array} \right) =  \det~(A-BD^{-1}C)(\det ~D)
 \en
 For our purposes we define the modified determinant $Det$:
  \eq
 \left[ Det~  \left( \begin{array}{cc}
   A_{\mu\nu} & B_{\mu h} \\
   C_{g\nu} & D_{g,h} \
 \end{array} \right) \right]^2= [{\det}_G ~(\Rcal_g
 D_{g,h})]^\dagger
 (\det M_{\mu\nu})^\dagger \det M_{\mu\nu}
 ~{\det}_G ~(\Rcal_g D_{g,h})
 \en
 with $M_{\mu\nu} \equiv A_{\mu\nu}-B_{\mu h}(D^{-1})^{h,g}
 C_{g,\nu}$ and
 \eq
  \det M_{\mu\nu} = \epsilon^{\mu_1...\mu_D} M_{\mu_1\nu_1}
  ...M_{\mu_D\nu_D}
  \epsilon^{\nu_1...\nu_D}
  \en

  \noi For the matrix $E_{A,B}$ the modified
 determinant reads
 \eq
  (Det~E_{A,B})^2 = [ {\det}_G ~(- \de_{g,h^{-1}} + F_{g,h})]^\dagger~
  (\det M_{\mu\nu})^\dagger \det M_{\mu\nu}
~{\det}_G ~(- \de_{g,h^{-1}} + F_{g,h}) \label{DetEAB}
 \en
 where
 \eqa
 & & M_{\mu\nu} \equiv g_{\mu\nu} + F_{\mu\nu} +  F_{\mu h} ~
  (H^{-1})^{h,g} ~ {\cal R}_{g^{-1}} F_{\nu g} \label{Mmunu}\\
 & & H_{g,h} \equiv - \de_{g,h^{-1}} + \Rcal_{g^{-1}}F_{g,h}
 \ena

 Next we prove that  $(Det~E_{AB})^2$ transforms covariantly.
 The matrix $H$ and its inverse $H^{-1}$ transform as
 \eq
H'_{g,h} = ({\cal R}_{g^{-1}} T ) ~ H_{g,h} ~ ({\cal R}_{h} T^{-1}
) \;\;\; \Rightarrow \;\;\; (H^{-1})^{h,g} = ({\cal R}_{h} T ) ~
(H^{-1})^{h,g} ~ ({\cal R}_{g^{-1}} T^{-1} )
 \en
 so that $M_{\mu\nu}$ transforms as $M'_{\mu\nu}=T M_{\mu\nu} T^{-1}$.
 As a consequence
 $(Det~E_{A,B})^2$ transforms covariantly
 \eq
 (Det~E'_{A,B})^2 = T ~(Det~E_{A,B})^2~ T^{-1}
 \en

 A real gauge invariant BI action on $M^D \times G$ can
 be constructed by taking twice the square root of
 $(Det~E_{AB})^2$:
 \eq
 A_{BI} = \int d^Dx ~ \sum_G Tr~ \sqrt{|(Det~E_{AB}|}
 \en

\noi {\bf Acknowledgements}
 \sk
 \noi We thank Chiara Pagani for useful
 comments.

 \noi The work of P.A. was supported by a Marie Curie Fellowship
 of the EC  programme IHP, contract number MCFI-2000-01982.
 The work of P.A. and L.C. was supported in part by EC under RTN
 project HPRN-CT-2000-00131.
 The work of A.P.I. was supported in part by the RFBR grant
 00-01-00299 and by the INFN-JINR(Dubna)
 program. P.A and A.P.I. are grateful to the
 Dipartimento di Fisica Teorica of Torino University,
 where work on this paper was started, and A.P.I also
 thanks the Max-Planck-Institut f\"{u}r Mathematik
 in Bonn, where he continued this work, for their kind hospitality
 and support.

\app{Hopf algebraic formulas for the differential calculus on
Fun(G)}


The $G$ group structure induces a Hopf algebra structure on
$Fun(G)$, with  coproduct $\D$,  coinverse $\kappa$ and counit
$\epsi$ defined by group multiplication, inverse and unit as:
 \eqa
& &\D (f) (g,g') = f(gg'),~~~\D:Fun(G) \rightarrow Fun(G) \otimes
Fun(G)\\ & &\kappa (f)(g) = f(g^{-1}),~~~~~~~\kappa: Fun(G)
\rightarrow Fun(G) \\ & &\epsi (f)=f(e),~~~~~~~~~~~~~~\epsi:
Fun(G) \rightarrow \Cb \ena In the first line we have used $Fun(G
\times G) \approx Fun(G) \otimes Fun(G)$ [indeed a basis for
functions on $G \times G$ is given by $x^{g_1} \otimes x^{g_2},
g_1,g_2 \in G$]. On the basis functions $x^g$ the costructures
take the form: \eq \D (x^g)=\sum_{h \in G} x^h \otimes x^{h^{-1}g}
\equiv x^g_{(1)} \otimes x^g_{(2)} ,~~\kappa
(x^g)=x^{g^{-1}},~~\epsi (x^g)=\de^g_e \label{cox}
 \en
 \sk
 \noi {\sl Left and right coactions on $\Ga$}
 \eq
 \Delta_L  (a d b)  = a_{(1)} \, b_{(1)} \otimes a_{(2)} \, d
b_{(2)} \; , \;\;\; \Delta_R  (a d b)  = a_{(1)} \, d b_{(1)}
\otimes a_{(2)} \, b_{(2)} \; .
 \en
 where $\Delta (a) \equiv a_{(1)} \otimes a_{(2)}$.
 \sk
 \noi{\sl  Left and right invariant forms}
 \eqa
& &\theta^g = \kappa (x^g_{(1)}) \, d \, x^g_{(2)} = \sum_h \, x^h
\, d \, x^{hg} =
 \sum_h \, x^{h \, g^{-1}} \, d \, x^{h} \\
 & & \zeta^g =  d(x^g_{(1)}) \, \kappa (x^g_{(2)}) = \sum_h \, (dx^{g
 h}) \, x^{h}= - \sum_h x^{gh}dx^h \; .
 \ena
 \noi{\sl Bimodule relations and $f^g_{~g'}$ functionals}
 \sk
 Definition of the $f^g_{~g'}$ functionals:
 \eq
 \theta^g a = (f^g_{~g'} * a)~\theta^{g'} \equiv (id \otimes
 f^g_{~g'}) \Delta (a) \theta^{g'}
  \en
 Applying the rule $\theta^g a= (\Rcal_g a) \theta^g$ (\ref{fthetacomm})
 to the basis functions $x^g$ yields
 \eq
 f^g_{~g'} = \de^g_{g'} g,~~~g\not=e,~g'\not=e
 \en
 where we denote as $g$ the functional on $Fun(G)$ dual
 to the basis function $x^g$, i.e. $g(x^{g'}) = \de^{g'}_g$
 \sk
 \noi {\sl Left and right coaction on $\Ga$}
 \sk
  Left and right coaction of $G$ on $\Ga$ are given by
   the mappings  $\DL:\Ga \rightarrow
Fun(G) \otimes \Ga$ and $\DR: \Ga \rightarrow \Ga \otimes Fun(G)$
that encode the information about all left or right translations:
 \eq
 \DL(a\rho b)=\D(a)\DL (\rho)\D(b),~~~\DL (db)=(id \otimes d) \D (b)
 ~~~ \forall a,b \in Fun(G),~\rho \in \Ga \label{leftco}
\en
\eq
 \DR(a\rho b)=\D(a)\DR (\rho)\D(b),~~~\DR (db)=(d \otimes id) \D (b)
 ~~~ \forall a,b \in Fun(G),~\rho \in \Ga \label{rightco}
\en
For example their action on the basic terms $x^{g_1}dx^{g_2} \in
\Ga$ is: \eq \DL (x^{g_1}dx^{g_2})=\D(x^{g_1})(id \otimes d) \D
(x^{g_2})= \sumonh x^{h}\otimes x^{h^{-1}g_1}
dx^{h^{-1}g_2}\label{DLxdx}
\en
\eq \DR (x^{g_1}dx^{g_2})=\D(x^{g_1})(d \otimes id) \D (x^{g_2})=
\sumonh x^{g_1 h} dx^{g_2 h} \otimes x^{h^{-1}} \label{DRxdx}
\en

Computing $\DL$ and $\DR$ on the basic differentials yields:
 \eqa
& & \DL (dx^{g_1}) \equiv (id \otimes d) (\D x^{g_1})= \sum_{h \in
G} x^h \otimes dx^{h^{-1}g_1}\\ & & \DR (dx^{g_1}) \equiv (d
\otimes id) (\D x^{g_1})= \sum_{h \in G} dx^h \otimes
x^{h^{-1}g_1} \ena \sk
 \noi {\sl Adjoint representation}
 \sk
 The adjoint representation matrix $M_{g'}^{~g}$ is defined by:
 \eq
 \DR (\theta^g)=\theta^{g'} \otimes M_{g'}^{~g}
 \en
 On the other hand computing the right coaction on $\theta^g$
 according to the formulas of the previous paragraph yields:
 \eq
 \DR (\theta^g)=\sum_h \theta^{hgh^{-1}} \otimes x^h=\sum_k
 \theta^k \otimes \sum_h \de^{hgh^{-1}}_k x^h
 \en
 so that
 \eq
 M_{g'}^{~g}=\sum_h \de^{hgh^{-1}}_{g'} x^h
 \en
 {\sl Braiding matrix $\La$}
 \eq
 \La^{gg'}_{~~k'k} = f^g_{~k}(M_{k'}^{~~g'})=\de^g_k
 \de^{gg'g^{-1}}_{k'},~~~g\not=e,~g'\not=e
 \en
 {\sl Hopf algebra structure of $\Omega$}
 \eq
\begin{array}{c}
\Delta(\theta^g) \equiv \Delta_L (\theta^g)  + \Delta_R (\theta^g)
= 1 \otimes \theta^g +  \sum_{h} \, \theta^{h g h^{-1}} \otimes
x^h \; , \\ \\ \epsilon (\theta^g ) =  0 \; , \;\;\;
\kappa(\theta^g) = - \sum_{h} \, \theta^{h^{-1} g h} \, x^h \; .
\label{Hopfomega}
\end{array}
 \en

\end{document}